\documentclass[epsf,useAMS,usenatbib,usegraphicx]{mn2e}
\usepackage{epsfig}
\usepackage{aas_macros} 
\usepackage{amsmath}
\usepackage{amssymb}
\usepackage{appendix}
\usepackage{booktabs}
\usepackage{lscape}
\usepackage{threeparttable,caption}
\usepackage{upgreek}
\usepackage{url}
\usepackage[usenames]{color}

\newcommand{\ion}[2]{{#1}\,{\sc #2}}
\newcommand{\specline}[3]{{#1}\,{\sc #2}\:{#3}}
\newcommand{\kepler}{\textit{Kepler}}

\title[Non-pulsating stars]{A search for non-pulsating, chemically normal stars in the $\delta$\,Scuti instability strip using \textit{Kepler} data}

\author[S. J. Murphy et al.] 
{Simon J. Murphy$^{1,2,\dagger}$, Timothy R. Bedding$^{1,2}$, Ewa Niemczura$^{3}$,  \and
Donald W. Kurtz$^{4}$ and Barry Smalley$^{5}$\\
$^1$Sydney Institute for Astronomy (SIfA), School of Physics, University of Sydney, Australia\\
$^2$Stellar Astrophysics Centre, Department of Physics and Astronomy, Aarhus University, 8000 Aarhus C, Denmark\\
$^{3}$Instytut Astronomiczny, Uniwersytet Wroc\l{}awski, Kopernika 11, 51-622 Wroc\l{}aw, Poland\\
$^{4}$Jeremiah Horrocks Institute, University of Central Lancashire, Preston, PR1 2HE, UK\\
$^{5}$Astrophysics Group, Keele University, Staffordshire, ST5 5BG, UK\\
\\
$^{\dagger}$email: murphy@physics.usyd.edu.au
}

\begin{document}

\maketitle 

\begin{abstract}
We identify stars in the $\delta$\,Sct instability strip that do not pulsate in p\:modes at the 50-$\upmu$mag limit, using \kepler\ data. Spectral classification and abundance analyses from high-resolution spectroscopy allow us to identify chemically peculiar stars, in which the absence of pulsations is not surprising. The remaining stars are chemically normal, yet they are not $\delta$\,Sct stars. Their lack of observed p\:modes cannot be explained through any known mechanism. However, they are mostly distributed around the edges of the $\delta$\,Sct instability strip, which allows for the possibility that they actually lie outside the strip once the uncertainties are taken into account.
We investigated the possibility that the non-pulsators inside the instability strip could be unresolved binary systems, having components that both lie outside the instability strip. If misinterpreted as single stars, we found that such binaries could generate temperature discrepancies of $\sim$300\,K --  larger than the spectroscopic uncertainties, and fully consistent with the observations.
After these considerations, there remains one chemically normal non-pulsator that lies in the middle of the instability strip. This star is a challenge to pulsation theory. However, its existence as the only known star of its kind indicates that such stars are rare. We conclude that the $\delta$\,Sct instability strip is pure, unless pulsation is shut down by diffusion or another mechanism, which could be interaction with a binary companion.
\end{abstract}



\section{Introduction}
\label{sec:intro}

Stars with spectral type A play an important role in stellar astrophysics. The thickness of surface convection zones decreases with increasing temperature, from the deep and efficient hydrogen convection zones in late F stars to the very weak helium convection zones of B stars \citep{cantielloetal2009}, making a transition in the A stars. Rotation has a major influence, broadening spectral lines and distorting the star. Indeed, the peak of the angular rotation frequency distribution occurs at A5, where stars commonly rotate at two-thirds of their break-up velocities \citep{royeretal2007}. Exceptions to this are chemically peculiar stars, whose rotation is orders of magnitude slower.

Pulsation is commonplace in late-A stars, where the classical instability strip intersects the main sequence and the $\delta$\,Sct stars are found. With the photometric precision afforded by the \kepler\ Mission, asteroseismology has enjoyed success with A stars, such as in the discovery of hundreds of $\delta$\,Sct--$\gamma$\,Dor hybrid stars \citep{grigahceneetal2010b,balona&dziembowski2011,beddingetal2014,vanreethetal2014}, the observation of tidally excited modes \citep{welshetal2011,thompsonetal2012}, and the determination of the surface-to-core rotation profile of a main sequence star for the first time \citep{kurtzetal2014}. However, with the exception of the roAp stars, pulsation and chemical peculiarity appear to be mutually exclusive. In these high-overtone pulsators, strong dipolar magnetic fields suppress the low-overtone pulsations seen in $\delta$\,Sct stars \citep{saio2005} and cause chemical abundance inhomogeneities \citep{stibbs1950}. Deciding whether or not pulsation and chemical peculiarity are mutually exclusive in the non-magnetic $\delta$\,Sct stars is important for our understanding of the driving mechanism, and is the focus of this paper.

Until now, attempts to decide this question have focussed on searching for pulsations in chemically peculiar Am stars. This method was practical, since classification spectra were obtainable in minutes, whereas time-series photometry required days of observations to demonstrate variability. Now, the situation is inverted. \kepler\ has provided light curves of thousands of $\delta$\,Sct stars, but spectra are lacking. Thus we invert the usual research question: instead of asking whether there are any chemically peculiar stars that pulsate, we search for chemically normal stars in the $\delta$\,Sct instability strip that do {\it not} pulsate.

The $\delta$\,Sct stars pulsate in low-order pressure (p) modes that are driven by the heat engine ($\kappa$) mechanism operating on the second ionisation zone of helium. Their periods range from a few hours to 15\,min \citep{holdsworthetal2014}, with the hotter and younger stars having the shorter periods. They overlap with the Am stars on the HR diagram, having spectral types around mid-to-late A and early F.

The chemically peculiar Am stars have abnormally strong absorption lines of most metals when compared to the strength of the Balmer lines (the classification of Am stars was described extensively by \citealt{gray&corbally2009}). The abundance anomalies in Am stars are thought to result from atomic diffusion: atoms with many absorption lines near flux maximum are radiatively levitated within the star, while those with few lines are easily ionised into argon-like configurations and sink \citep{baglinetal1973}. In the absence of efficient mixing mechanisms, most transition metals become overabundant near the surface. Turbulence opposes the elemental segregation, but is negligible in slow rotators, hence the Am stars are observed with $v\sin i \lesssim 100$\,km\,s$^{-1}$ \citep{slettebak1954,slettebak1955}. The settled elements include Ca, Sc and He. The first two allow the Am stars to be distinguished from generic metal-rich stars, while the latter has important consequences for pulsational driving.

For nearly two decades, it was thought that Am stars did not pulsate. \citet{breger1970} first observed the apparent non-overlap between Am peculiarities and $\delta$\,Sct pulsation. The observations supported theory: gravitational settling of helium in slow rotators was thought to inhibit $\delta$\,Sct pulsations \citep{baglinetal1973}. Several counterexamples to Breger's observations were put forward, but each was dismissed \citep{kurtzetal1976}, until the classical Am star HD\,1097 was found to pulsate \citep{kurtz1989}. Dozens of pulsating Am stars are now known \citep{smalleyetal2011}, but it remains the case that the majority of Am stars do not pulsate. Attempts to explain pulsating Am stars have involved modelling an Fe-peak opacity bump at $\sim$200\,000\,K \citep{richeretal2000} and the incorporation of additional mixing mechanisms that downscale the abundance anomalies in Am star models to be more in line with observations \citep{turcotteetal2000, theadoetal2009}.

The existence of pulsations in some Am stars led \citet{catanzaro&balona2012} to describe `the helium problem in Am stars' in their eponymous paper. Here, we show that as far as chemically normal stars are concerned, the paucity of non-pulsators in the $\delta$\,Sct instability strip is in accordance with theory.


\section{Method}

\subsection{Overview}
Our method was to select non-$\delta$\,Sct stars observed by \kepler\ and to determine whether they lie in the $\delta$\,Sct instability strip. The selection was made on the absence of p\:modes -- low frequency variability from g\:modes or rotation were ignored in the selection, as further discussed in \S\,\ref{ssec:targets}. Note that the non-$\delta$\,Sct stars lying outside of the instability strip do not pulsate because they are either too hot, such that the \ion{He}{ii} partial ionisation zone lies too close to the surface to drive pulsation efficiently, or too cool, such that the surface convection zone is so deep as to dampen the pulsation completely.

\begin{figure*}
\begin{center}
\includegraphics[width=0.96\textwidth]{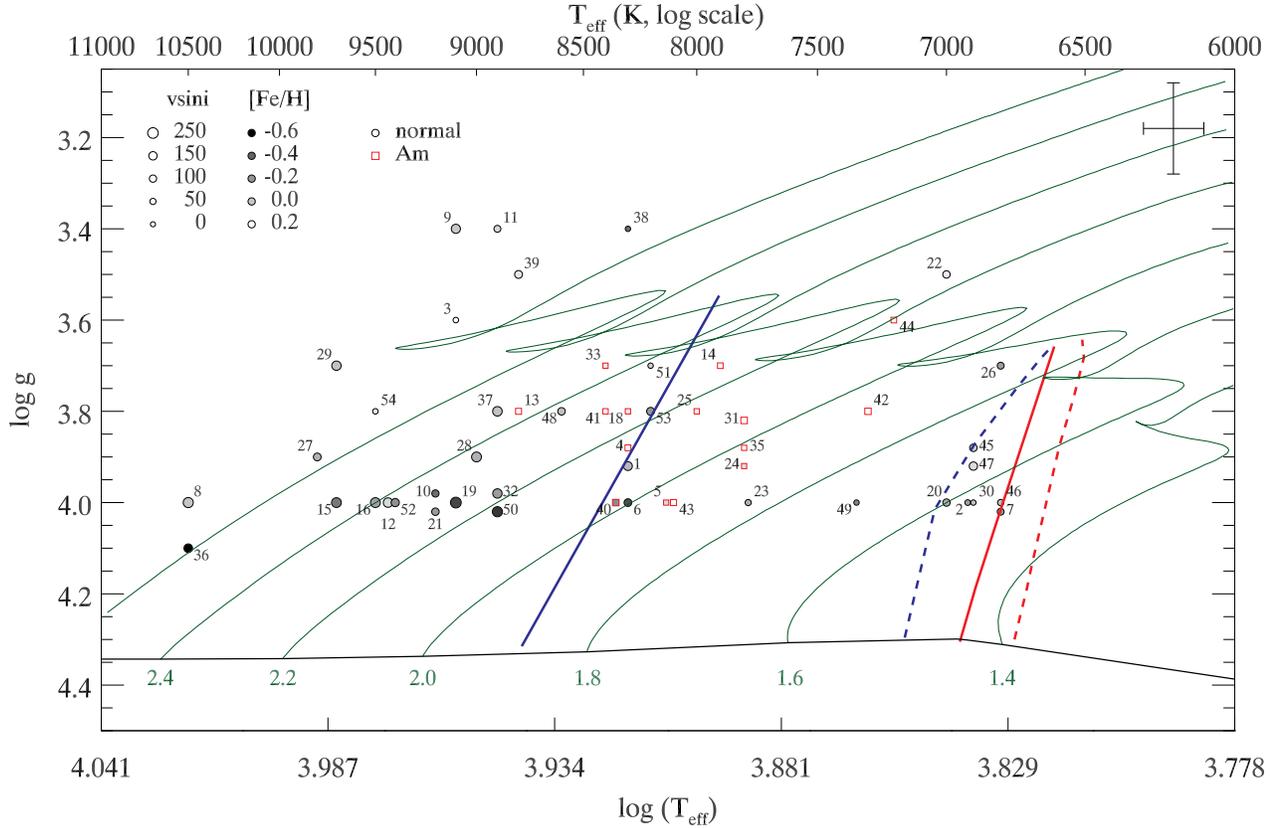}
\caption{$T_{\rm eff}$-$\log g$ diagram of 54 non-$\delta$\,Sct stars. {\it Green lines} are evolutionary tracks computed with time-dependent convection models. The corresponding masses in M$_{\odot}$ are written beneath the ZAMS ({\it black line}). The blue and red edges of the $\delta$\,Sct instability strip are represented by {\it solid blue and red lines}. Similarly, the $\gamma$\,Dor instability strip is represented by {\it dashed lines}. The two instability strips cover the full main-sequence band, hence for a main-sequence star pulsational instability is mostly determined by the effective temperature. We plot stellar positions of our sample with spectroscopic $T_{\rm eff}$ and $\log g$ values from \citet[][in review]{niemczuraetal2015}. Symbol {\it fill colour} indicates [Fe/H] and symbol {\it size} is proportional to $v\sin i$. {\it Red squares} are Am stars, while {\it black circles} are chemically normal stars. Typical error bars are represented by the point in the top right. The small numbers by each symbol indicate the Star\# from Table\:\ref{tab:parameters}, as discussed in \S\,\ref{ssec:atmospheric-parameters}.}
\label{fig:specteff}
\end{center}
\end{figure*}

Determining whether a star lies in the instability strip requires an accurate measurement of effective temperature but only an approximate surface gravity. As shown in Fig.\,\ref{fig:specteff}, the $\delta$\,Sct instability strip occupies the full main sequence band, and even some pre- and post-main sequence $\delta$\,Sct stars are known (see, e.g., \citealt{lenzetal2010}; \citealt{casey2011}; \citealt{zwintzetal2014}), reducing the importance of a precise surface gravity. Furthermore, metallicity affects both evolutionary tracks and instability strip boundaries of $\delta$\,Sct stars, but the stars in our sample had metallicities typically no more than $\pm0.5$\,dex from solar. Since the former effect is the greater of the two, but is reflected in the observed $T_{\rm eff}$ of the star, we focussed our efforts on constraining the stellar $T_{\rm eff}$.

Once the $T_{\rm eff}$ of each star was determined, a reasonable approximation to $\log g$ allowed us to place the star on a $T_{\rm eff}$-$\log g$ diagram and evaluate its position relative to the instability strip. We used instability strip boundaries for stars of solar metallicity based on \citet{dupretetal2005b}, but calculated the $\delta$\,Sct boundaries for all unstable modes that have $\ell \leq 2$. Our boundaries are thus wider than conservative ones computed for radial modes only. We used time-dependent convection with a mixing length theory (MLT) parameter $\alpha_{\rm MLT} = 1.8$. This is the number of pressure scale heights travelled by a parcel of convective gas. We used evolutionary tracks computed with the same physics \citep{grigahceneetal2005}.

\subsection{Target selection}
\label{ssec:targets}
We analysed the pulsational content of \kepler\ light curves of all bright ($V\lesssim10$) stars whose $T_{\rm eff}$ values in the Kepler Input Catalogue (KIC, \citealt{brownetal2011}) indicated they were in or near the $\delta$\,Sct instability strip. We put the `non-pulsator' limit at 50\,$\upmu$mag, which corresponds to the maximum noise levels expected in Fourier transforms of \textit{Kepler} light curves (see \citealt{murphy2014}, Ch.\,1.3), and applied it to the p-mode frequency regime ($\sim 5$ to $50$\,d$^{-1}$). \citet{murphy2014} found that 44\:per\:cent of stars in the $\delta$\,Sct instability strip are non-$\delta$\,Sct stars by this definition. There is no theoretical exclusion between Am stars and gravity-mode (g-mode) oscillations, such as those seen in $\gamma$\,Dor stars, or between g\:modes and p\:modes in the same star. Indeed, $\gamma$\,Dor--$\delta$\,Sct hybrids are common \citep{hareter2012}. Thus we did not consider g\:modes in our selection, and so any peaks below 5\,d$^{-1}$ were not considered, though we do discuss these in \S\,\ref{sec:discussion}. We did not explicitly look for roAp oscillations, which can be found at late-A spectral types, since Ap stars would be detected in spectroscopic investigations and later filtered out.

In our Fourier analysis we used \kepler\ long-cadence (LC) data processed with the msMAP pipeline (see, e.g., \citealt{stumpeetal2012}, \citealt{smithetal2012}). The LC data have a 29.45-min cadence, and a corresponding Nyquist frequency of 24.4\,d$^{-1}$. However, our analysis is not hindered by Nyquist ambiguity. If any of our targets were p-mode oscillators, we would see either their real pulsation frequencies or their Nyquist aliases in the range 0--24.4\,d$^{-1}$. \citet{murphyetal2012b} showed how the periodic modulation of the sampling times of \kepler\ data, due to the orbital motion of the satellite around the Sun, alleviates Nyquist ambiguity entirely, and the real pulsation frequency can be recovered. Hence LC data are adequate for our study.

\subsection{Temperature Determinations}
\label{ssec:temperatures}

Further observations of the non-$\delta$\,Sct stars were necessary to confirm they lie in the instability strip because KIC temperatures are inaccurate for hot ($T_{\rm eff} \gtrsim 6500$\,K) stars \citep{pinsonneaultetal2012}. For this paper, we made additional observations with the HERMES spectrograph \citep{raskinetal2011} at the 1.2-m Mercator telescope on La Palma. The resulting spectra were collated with spectra of A and F stars from other projects that have been analysed by \citet[][in review]{niemczuraetal2015}. We use their atmospheric parameters directly. We also determined temperatures from spectral energy distribution (SED) fitting, and from the revised KIC \citep{huberetal2014}. These two additional methods of determining temperature were used to verify the spectroscopic results. Our total sample consisted of 54 non-$\delta$\,Sct stars that have HERMES spectra.

\subsubsection{High-resolution spectroscopy}

The Balmer lines of hydrogen and lines of \ion{Fe}{i} are sensitive to temperature in the A stars. This sensitivity is utilised during spectroscopic analysis to determine the effective temperature to a precision of $\sim100$\,K. We used temperatures obtained in this manner by \citet{niemczuraetal2015}, along with [Fe/H], $\log g$ and $v\sin i$.

\subsubsection{Spectral Energy Distributions}

An independent measure of temperature comes from SEDs. The majority of the SED temperatures used here also come from \citet{niemczuraetal2015}, with the exception of six double-lined spectroscopic binary systems (SB2s) that were not reported there. For those, we followed the same method as \citet{niemczuraetal2015}, that is, we obtained approximate SEDs from the following available photometry: 2MASS \citep{skrutskieetal2006}, Tycho $B$ and $V$ magnitudes \citep{hoegetal1997}, USNO-B1 $R$ magnitudes \citep{monetetal2003}, TASS $I$ magnitudes \citep{droegeetal2006}, Geneva photometry \citep{rufener1999}, and CMC14 $r'$ magnitudes \citep{evansetal2002}.

Effective temperatures were determined by fitting \citet{kurucz1993a} model fluxes to the de-reddened SEDs. The model fluxes were convolved with photometric filter response functions. A weighted Levenberg-Marquardt nonlinear least-squares fitting procedure was used to find the solution that minimized the difference between the observed and model fluxes. Since $\log g$ and metallicity [M/H] are poorly constrained by our SEDs, we fixed $\log g = 4.0$ and [M/H]$= 0.0$\,dex for all the fits. The uncertainties in $T_{\rm eff}$ include the formal least-squares error and adopted uncertainties in $E(B-V)$ of $\pm0.02$\,mag, $\log g$ of $\pm0.5$\,dex and [M/H] of $\pm0.5$\,dex added in quadrature.

\subsubsection{Revised KIC temperatures}

\citet{huberetal2014} incorporated photometry and spectroscopy in their revision of KIC parameters. They provided $T_{\rm eff}$, $\log g$ and [Fe/H] along with uncertainties for all A stars observed by \kepler.

\subsection{Collation of atmospheric parameters}
\label{ssec:atmospheric-parameters}

We have collated atmospheric parameters for the 54 targets into Table\:\ref{tab:parameters}. These are all non-$\delta$\,Sct stars down to our threshold of 50$\upmu$mag, but some are $\gamma$\,Dor stars. We also give the spectral types, which were determined by smoothing the HERMES spectra to classification resolution (1.8\,\AA\ per 2\,px) and comparing against a series of high- and low-$v\sin i$ standards collected by R.O.~Gray and one of us (SJM; Murphy et al., in prep.). The spectral types indicate which stars are Am.

\begin{table*}
\begin{threeparttable}
\caption{Parameters for the targets. Column headers use the same conventions as in the text, with new additions of: star\# for the ID of each star; Spec. to denote values from high resolution spectroscopy; and Sp. T. for spectral type. A superscript $^B$ after a star number indicates it is an SB2 system. An asterisk in the final column signifies a note to the spectral type, as determined in \citet{niemczuraetal2015}. Spectroscopic uncertainties on $T_{\rm eff}$ and $\log g$ are $100$\,K and 0.1\,dex, respectively. [Fe/H] uncertainties range from 0.07 to 0.15\,dex. $v\sin i$ uncertainties are typically $\sim5$\:per\:cent or 2\,km\,s$^{-1}$ -- whichever the greater.}
\begin{tabular}{ccrcrrrcl}
\toprule
	Star \#	&	SIMBAD name	&	KIC ID	&$	Kp	$&	Spec. $T_{\rm eff}$	&	Spec. $\log g$	&	$v \sin i$	&	Spec. [Fe/H]	&	Sp. T.			\\
		&		&		&	mag	&	K	&	(cgs)	&	km\,s$^{-1}$	&	dex	&				\\
\midrule
$	1	$&	BD+36 3554	&$	1294756	$&$	9.1	$&$	8300	$&$	3.9	$&$	170	$&$	-0.06	$&	A3 IV			\\
$	2^B	$&	BD+37 3340	&$	2695344	$&$	9.6	$&$	6921	$&$	4.0	$&$	30	$&$	-0.16	$&	F2 V			\\
$	3	$&	BD+38 3558	&$	3231985	$&$	9.3	$&$	9100	$&$	3.6	$&$	22	$&$	\phantom{-}0.29	$&	A4 IV	Ca weak (A3)	*	\\
$	4	$&	BD+38 3565	&$	3441230	$&$	9.9	$&$	8300	$&$	3.9	$&$	36	$&$	\phantom{-}0.30	$&	Am kA2hA5mF0 V		\\
$	5^B	$&	BD+38 3679	&$	3656913	$&$	9.9	$&$	8132	$&$	4.0	$&$	10	$&$	\phantom{-}0.07	$&	Am kA2hA6mF0 IV		\\
$	6	$&	BD+38 3594	&$	3851151	$&$	9.8	$&$	8300	$&$	4.0	$&$	119	$&$	-0.47	$&	A3 V			\\
$	7	$&	BD+38 3468	&$	3942392	$&$	9.9	$&$	6800	$&$	4.0	$&$	82	$&$	-0.22	$&	F2 Vs			\\
$	8	$&	BD+38 3580	&$	4056136	$&$	9.6	$&$	10500	$&$	4.0	$&$	227	$&$	-0.01	$&	B9 IV-Vnn			\\
$	9	$&	TYC 3139-1375-1	&$	4572373	$&$	9.9	$&$	9100	$&$	3.4	$&$	184	$&$	\phantom{-}0.00	$&	A3 Van	(wk met A2)	*	\\
$	10	$&	HD 226221	&$	4681323	$&$	9.1	$&$	9200	$&$	4.0	$&$	99	$&$	-0.36	$&	A1 IV$^-$s		*	\\
$	11	$&	BD+39 3743	&$	4831769	$&$	9.6	$&$	8900	$&$	3.4	$&$	84	$&$	\phantom{-}0.06	$&	A3 Va			\\
$	12^B	$&	HD 182442	&$	4832225	$&$	9.1	$&$	9436	$&$	4.0	$&$	200	$&$	\phantom{-}0.07	$&	B9.5 V			\\
$	13	$&	HD 225365	&$	5199439	$&$	9.6	$&$	8800	$&$	3.8	$&$	40	$&$	\phantom{-}0.27	$&	Am: kA3hA5mA5 (IV)s		\\
$	14	$&	HD 225410	&$	5200084	$&$	9.2	$&$	7900	$&$	3.7	$&$	27	$&$	\phantom{-}0.43	$&	Am kA3hA6mF1 (IV)	*	\\
$	15	$&	HD 225785	&$	5294231	$&$	9.8	$&$	9700	$&$	4.0	$&$	213	$&$	-0.32	$&	A0 IVn			\\
$	16	$&	BD+40 3639	&$	5524045	$&$	9.5	$&$	9500	$&$	4.0	$&$	215	$&$	-0.15	$&	A0.5 Va$^+$			\\
$	17	$&	TYC 3125-1342-1	&$	5525210	$&$	9.9	$&$	6800	$&$	4.1	$&$	90	$&$	\phantom{-0.13}	$&	F1\,V **	\\
$	18	$&	HD 225463	&$	5633448	$&$	9.0	$&$	8300	$&$	3.8	$&$	13	$&$	\phantom{-}0.13	$&	Am: kA3hA4mA7 V	*	\\
$	19	$&	HD 182192	&$	5786771	$&$	9.1	$&$	9100	$&$	4.0	$&$	257	$&$	-0.50	$&	A0.5nn:		*	\\
$	20	$&	HD 181654	&$	5954264	$&$	8.2	$&$	7000	$&$	4.0	$&$	105	$&$	-0.11	$&	F1 IV-V			\\
$	21	$&	HD 177061	&$	6106152	$&$	8.1	$&$	9200	$&$	4.0	$&$	108	$&$	-0.17	$&	A3 IV	wk met (A1)		\\
$	22	$&	HD 184521	&$	6128236	$&$	8.9	$&$	7000	$&$	3.5	$&$	106	$&$	\phantom{-}0.14	$&	F0 IV$^+$			\\
$	23^B	$&	HD 179337	&$	6192566	$&$	9.4	$&$	7783	$&$	4.0	$&$	50	$&$	-0.08	$&	A6 IV-V			\\
$	24	$&	TYC 3143-1942-1	&$	6292398	$&$	9.8	$&$	7800	$&$	3.9	$&$	8	$&$	\phantom{-}0.09	$&	Am: kA3hA7mA7 V		\\
$	25	$&	BD+41 3418	&$	6292925	$&$	9.7	$&$	8000	$&$	3.8	$&$	15	$&$	\phantom{-}0.16	$&	Am: kA2.5hA3mA7 (IV)		\\
$	26	$&	HD 225711	&$	6380579	$&$	9.8	$&$	6800	$&$	3.7	$&$	83	$&$	-0.17	$&	F2 Vs			\\
$	27	$&	HD 185265	&$	6450107	$&$	7.6	$&$	9800	$&$	3.9	$&$	121	$&$	-0.08	$&	A1 IV-s			\\
$	28	$&	HD 177328	&$	7345479	$&$	7.9	$&$	9000	$&$	3.9	$&$	215	$&$	-0.09	$&	A2 Vnn			\\
$	29	$&	HD 184024	&$	7530366	$&$	8.4	$&$	9700	$&$	3.7	$&$	193	$&$	-0.02	$&	A0.5 IVnn			\\
$	30	$&	TYC 3130-497-1	&$	7661054	$&$	9.9	$&$	6900	$&$	4.0	$&$	16	$&$	-0.02	$&	F2.5 V			\\
$	31	$&	HD 186995	&$	7767565	$&$	9.3	$&$	7800	$&$	3.8	$&$	65	$&$	\phantom{-}0.41	$&	Am kA5hA7mF1 IV\\
$	32	$&	HD 182952	&$	8027456	$&$	9.7	$&$	8900	$&$	4.0	$&$	194	$&$	-0.17	$&	A1 V			\\
$	33	$&	HD 187091	&$	8112039	$&$	8.4	$&$	8400	$&$	3.7	$&$	9	$&$	\phantom{-}0.12	$&	Am: kA3hA5mA5 (IV)s		\\
$	34	$&	HD 173978	&$	8211500	$&$	8.1	$&$	7800	$&$	3.8	$&$	93	$&$	\phantom{-}0.15	$&	A5 IV **			\\
$	35	$&	HD 188911	&$	8323104	$&$	9.7	$&$	7800	$&$	3.9	$&$	10	$&$	\phantom{-}0.20	$&	Am kA2.5hA6mA7 (IV)\\
$	36	$&	HD 177152	&$	8351193	$&$	7.6	$&$	10500	$&$	4.1	$&$	162	$&$	-1.27	$&	B9.5:		*	\\
$	37	$&	HD 184023	&$	8367661	$&$	8.7	$&$	8900	$&$	3.8	$&$	201	$&$	-0.02	$&	A2 IVn			\\
$	38	$&	HD 188539	&$	8386982	$&$	9.8	$&$	8300	$&$	3.4	$&$	14	$&$	-0.37	$&	A4 IV/V:		*	\\
$	39	$&	HD 181598	&$	8489712	$&$	8.6	$&$	8800	$&$	3.5	$&$	126	$&$	\phantom{-}0.12	$&	A2 IVs			\\
$	40^B	$&	HD 184482	&$	8692626	$&$	8.3	$&$	8354	$&$	4.0	$&$	65	$&$	-0.24	$&	Am: kA2hA4mA6\,(IV)	*	\\
$	41	$&	HD 187254	&$	8703413	$&$	8.7	$&$	8400	$&$	3.8	$&$	14	$&$	\phantom{-}0.59	$&	Am kA3hA5mA9 (IV)s		\\
$	42	$&	HD 190165	&$	9117875	$&$	7.5	$&$	7300	$&$	3.8	$&$	61	$&$	\phantom{-}0.51	$&	Am kA3hF0.5mF3 (III)	*	\\
$	43	$&	HD 180239	&$	9147002	$&$	9.9	$&$	8100	$&$	4.0	$&$	42	$&$	\phantom{-}0.35	$&	Am kA3hA5mF3 (IV)	 *	\\
$	44	$&	HD 176843	&$	9204718	$&$	8.8	$&$	7200	$&$	3.6	$&$	28	$&$	\phantom{-}0.21	$&	Am kA3hA9mF1 V		\\
$	45	$&	HD 185329	&$	9286638	$&$	7.3	$&$	6900	$&$	3.9	$&$	148	$&$	\phantom{-}0.08	$&	F2 V	wk met (F0)		\\
$	46	$&	HD 188713	&$	9300946	$&$	9.9	$&$	6800	$&$	4.0	$&$	58	$&$	-0.04	$&	F2 Vs		*	\\
$	47	$&	BD+45 2978	&$	9419182	$&$	9.2	$&$	6900	$&$	3.9	$&$	110	$&$	\phantom{-}0.02	$&	F2 V			\\
$	48	$&	HD 178508	&$	9699848	$&$	9.1	$&$	8600	$&$	3.8	$&$	110	$&$	-0.03	$&	A2.5 Vn			\\
$	49^B	$&	TYC 3561-609-1	&$	10026614	$&$	9.8	$&$	7345	$&$	4.0	$&$	15	$&$	-0.26	$&	F4 V:			\\
$	50	$&	HD 179069	&$	10263800	$&$	9.8	$&$	8900	$&$	4.0	$&$	230	$&$	-0.54	$&	A0.5 Vn			\\
$	51	$&	BD+47 2777	&$	10721930	$&$	9.7	$&$	8200	$&$	3.7	$&$	16	$&$	\phantom{-}0.03	$&	A5 Vs met str (A6)		*	\\
$	52	$&	HD 183257	&$	11189959	$&$	8.2	$&$	9400	$&$	4.0	$&$	143	$&$	-0.22	$&	A1 Va$^+$ n			\\
$	53	$&	BD+49 3007	&$	11506607	$&$	9.7	$&$	8200	$&$	3.8	$&$	129	$&$	-0.19	$&	A3 V			\\
$	54	$&	HD 179617	&$	12153021	$&$	8.7	$&$	9500	$&$	3.8	$&$	20	$&$	\phantom{-}0.13	$&	A2 Va$^+$s			\\
\bottomrule
\end{tabular}
\begin{tablenotes}
\item [*] Notes on spectral type by star number are given in Appendix\,\ref{ap:spectraltype}
\item[**] Stars \#17 and \#34 were removed from the analysis hereafter due to their accidental inclusion in an early draft on the stars with HERMES spectra \citep{niemczuraetal2015}.
\end{tablenotes}
\label{tab:parameters}
\end{threeparttable}
\end{table*}

We assume that stars that appear chemically normal in spectroscopy have homogeneous abundances. Rapidly rotating A stars are well mixed, so the metallicity measured in the line forming region by spectroscopy is equivalent to that in deeper regions as probed by pulsation. An exception would be the $\lambda$\,Boo stars, which often rotate rapidly and have surface peculiarities that are presumably replenished from accretion of circumstellar material \citep{venn&lambert1990, king1994}, or that are short-lived \citep{turcotte&charbonneau1993, kamp&paunzen2002}. With asteroseismology it is possible to distinguish between surface peculiarities, like the selective metal weakness characteristic of $\lambda$\,Boo stars, or global metal weakness common in Pop.\,II stars, as was done by \citet{murphyetal2013a}. As yet, no $\delta$\,Sct star has been shown to have a peculiar interior with a normal abundance pattern at the surface.

In Fig.\,\ref{fig:specteff} we show the locations of stars in the $T_{\rm eff}$--$\log g$ diagram, using values from high-resolution spectroscopy. We also show [Fe/H] and $v\sin i$. Some stars overlapped on this plot due to the quantised $T_{\rm eff}$ and $\log g$ intervals used in the spectroscopic analysis and so they have been offset by up to $\pm$0.02\,dex in $\log g$ for clarity -- a value smaller than the observational uncertainty.

It is clear that the majority of our non-$\delta$\,Sct stars are hotter than the blue edge of the instability strip, although none is cooler than the red edge. About a third of the sample lie within the $\delta$\,Sct instability strip. Many of these stars are Am stars, which is not surprising because it agrees with the idea that Am stars have insufficient helium in the driving region to excite pulsation. However, there are several chemically normal stars inside the instability strip that are not pulsating, contrary to expectation.


\section{Discussion}
\label{sec:discussion}

\begin{figure*}
\begin{center}
\includegraphics[width=0.92\textwidth]{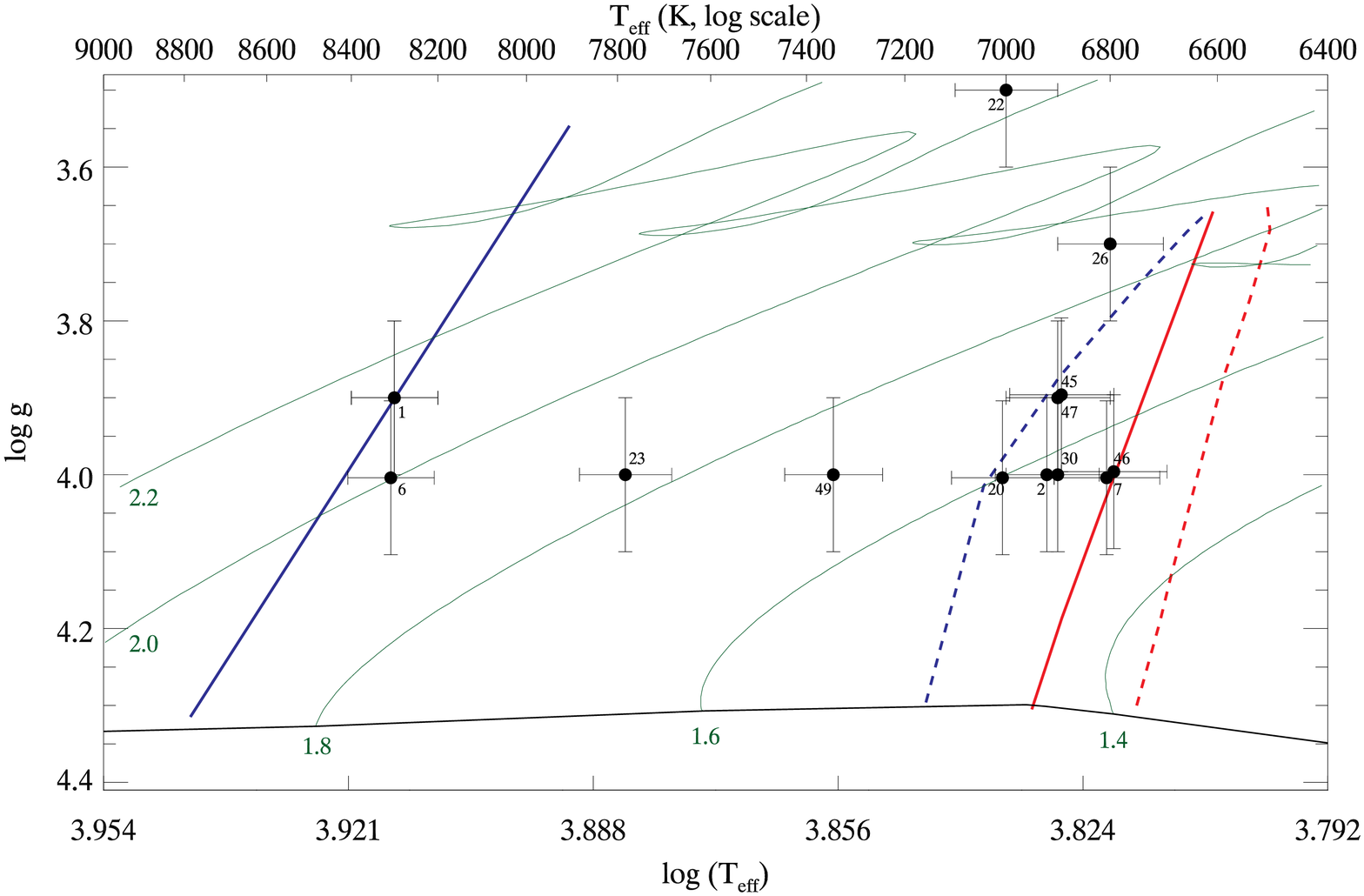}
\caption{Positions of chemically normal, non-$\delta$\,Sct stars with $1\sigma$ error bars. As in Fig.\,\ref{fig:specteff}, {\it solid blue and red lines} are the blue and red edges of the $\delta$\,Sct instability strip, while {\it dashed lines} indicate the $\gamma$\,Dor instability strip. {\it Green lines} are evolutionary tracks, with masses in M$_{\odot}$ written beneath the ZAMS ({\it black}). The non-pulsators generally lie near the edges of the $\delta$\,Sct instability strip, with exceptions discussed in the text.}
\label{fig:zoom}
\end{center}
\end{figure*}

We selected stars for further investigation if they were located within the instability strip boundaries (Fig.\,\ref{fig:specteff}) and were not Am stars. A total of thirteen stars met this criterion. Neither their spectral types, nor their abundances \citep{niemczuraetal2015} indicate chemical peculiarity. They are listed in Table\:\ref{tab:selected}, and their locations in the $T_{\rm eff}$--$\log g$ diagram are shown in Fig.\,\ref{fig:zoom}.

\begin{table*}
\begin{threeparttable}
\caption{Atmospheric parameters of stars that lie within the instability strip. Superscripted `B' characters after the Star\# indicate SB2 systems. SED $T_{\rm eff}$ values are shown, which agree with spectroscopic $T_{\rm eff}$ values to 1$\sigma$. Hence we use these for the SB2 systems where no spectroscopic $T_{\rm eff}$ is yet available, and we assume $\log g = 4.0$. For these, $v\sin i$ values are only approximate and are based on the primary star.}
\renewcommand{\arraystretch}{1.4}
\begin{tabular}{lrcccr}
\toprule
	Star\#	&	KIC ID	&	Spec. $T_{\rm eff}$			&	Spec. $\log g$		&	SED $T_{\rm eff}$ 	&	$v\sin i$ \\
		&		&	K			&	(cgs)			&	K			&	km\,s$^{-1}$	\\
\midrule
$	\phantom{1}1	$&$	1294756	$&$	8300	\pm	100	$&$	3.9	\pm	0.1	$&$	8411	\pm	434	$&$	170	$\\
$	\phantom{1}2^B	$&$	2695344	$&$		---		$&$		---		$&$	6910	\pm	150	$&$	30	$\\
$	\phantom{1}6	$&$	3851151	$&$	8300	\pm	100	$&$	4.0	\pm	0.1	$&$	8275	\pm	418	$&$	119	$\\
$	\phantom{1}7	$&$	3942392	$&$	6800	\pm	100	$&$	4.0	\pm	0.1	$&$	6935	\pm	306	$&$	82	$\\
$	20^{\dagger}	$&$	5954264	$&$	7000	\pm	100	$&$	4.0	\pm	0.1	$&$	7057	\pm	311	$&$	105	$\\
$	22	$&$	6128236	$&$	7000	\pm	100	$&$	3.5	\pm	0.1	$&$	7088	\pm	317	$&$	106	$\\
$	23^B	$&$	6192566	$&$		---		$&$		---		$&$	7850	\pm	210	$&$	50	$\\
$	26	$&$	6380579	$&$	6800	\pm	100	$&$	3.9	\pm	0.1	$&$	6747	\pm	313	$&$	83	$\\
$	30	$&$	7661054	$&$	6900	\pm	100	$&$	4.0	\pm	0.1	$&$	6777	\pm	365	$&$	16	$\\
$	45^{\diamond}	$&$	9286638	$&$	6900	\pm	100	$&$	3.9	\pm	0.1	$&$	6926	\pm	298	$&$	148	$\\
$	46	$&$	9300946	$&$	6800	\pm	100	$&$	4.0	\pm	0.1	$&$	6912	\pm	374	$&$	58	$\\
$	47	$&$	9419182	$&$	6900	\pm	100	$&$	3.9	\pm	0.1	$&$	6929	\pm	302	$&$	110	$\\
$	49^B	$&$	10026614	$&$		---		$&$		---		$&$	6610	\pm	130	$&$	15	$\\
\bottomrule
\end{tabular}
\begin{tablenotes}
\small
\item[$\dagger$] $T_{\rm eff}$ values are in agreement with published $T_{\rm eff} = 6935$\,K \citep{casagrandeetal2011}.
\item[$\diamond$] $T_{\rm eff}$ values are in agreement with published $T_{\rm eff} = 6987$\,K \citep{casagrandeetal2011}.
\end{tablenotes}
\label{tab:selected}
\end{threeparttable}
\end{table*}

\subsection{Non-pulsators inside the $\delta$\,Sct instability strip}

In Fig.\,\ref{fig:zoom} the non-pulsators generally lie near the edges of the $\delta$\,Sct instability strip. The three exceptions are \#22, \#23, and \#49. The remaining stars are at most $\sim$200\,K (2$\sigma$) from the instability strip boundaries, so their placement inside the $\delta$\,Sct instability strip is still uncertain. Stars \#23 and \#49 are double-lined spectroscopic binaries (SB2 systems), for which spectroscopic temperatures are still to be determined. In Fig.\,\ref{fig:zoom} they are shown using their SED temperatures, which do not resolve the components and may therefore be incorrect.
Star \#22 is not known to be a binary. Evolutionary tracks place it in the hydrogen-shell-burning stage of evolution, where asteroseismic models predict a large number of pulsation modes to be excited \citep{lenzetal2010}, yet it shows none.

The low $\log g$ of Star \#22 raises suspicions that it might be a member of an undetected binary system. Binaries can easily give discrepant temperatures if treated as single stars, even for slow rotators where the lines are easier to disentangle. KOI-54 (not in our sample) is a prime example. It is a highly eccentric binary system, consisting of two slowly rotating stars ($v\sin i = 7.5$ and 4.5\,km\,s$^{-1}$, \citealt{welshetal2011}). Both components lie just blueward of the blue edge of the $\delta$\,Sct instability strip, having $T_{\rm eff} = 8500$ and 8800\,K. When \citet{niemczuraetal2015} observed this system it was at conjunction, and analysis of the single-lined spectrum yielded $T_{\rm eff} = 8400$\,K.  Do similar arguments apply for Star \#22? This star is a moderate rotator with $v\sin i \sim 100$\,km\,s$^{-1}$, for which small amounts of spectral contamination could still go undetected but the derived $T_{\rm eff}$ would need to be wrong by around 500\,K (i.e. 5\,$\sigma$, for the 100-K spectroscopic uncertainties) to shift Star \#22 outside of the instability strip. Appendix\,\ref{ap:binaries} shows this is unlikely, if Star \#22 were hypothesized to be a long-period ($>20$-d) binary.

SB2 systems are more obvious when the orbital period is shorter and the orbital velocities are greater. However, there is a lower-limit on orbital periods before tidal effects come into play. Am stars are commonly found in close binary systems\footnote{The binary fraction of Am stars is at least 60--65\:per\:cent \citep{debernardietal2000, carquillat&prieur2007, smalleyetal2014}, compared to just 35\:per\:cent for the A stars as a whole \citep{abt2009}. See \citet{murphy2014} for a review.}. It is thought that the presence of a close binary companion provides the tidal braking necessary to slow rotation. Atomic diffusion can then operate relatively uninhibited by mixing processes associated with rotation, and the star becomes an Am star. Spectra showing obvious Am peculiarities were excluded from our analysis, thus short-period binaries would not lead to chemically normal non-pulsators inside the instability strip, and our justification of a 20-d minimum orbital period in the simulations is justified.

None the less, it is noteworthy that the known SB2s in our sample do have low $v\sin i$. It remains puzzling that they are not Am stars. We might postulate that these stars are rapid rotators seen at low inclination, but confirmation will require detailed analyses of the SB2 systems and additional radial velocity follow-up.

\begin{figure*}
\begin{center}
\includegraphics[width=0.38\textwidth]{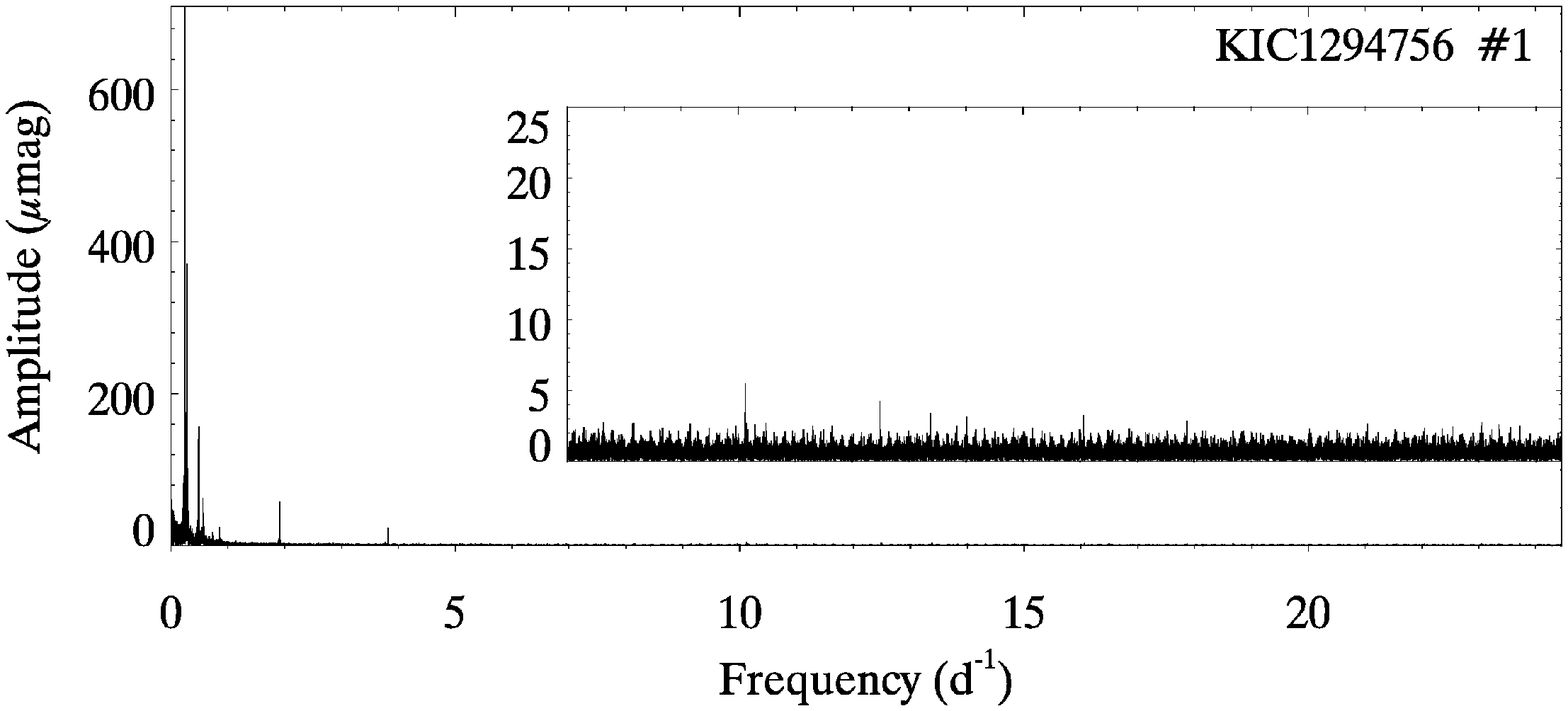}
\includegraphics[width=0.38\textwidth]{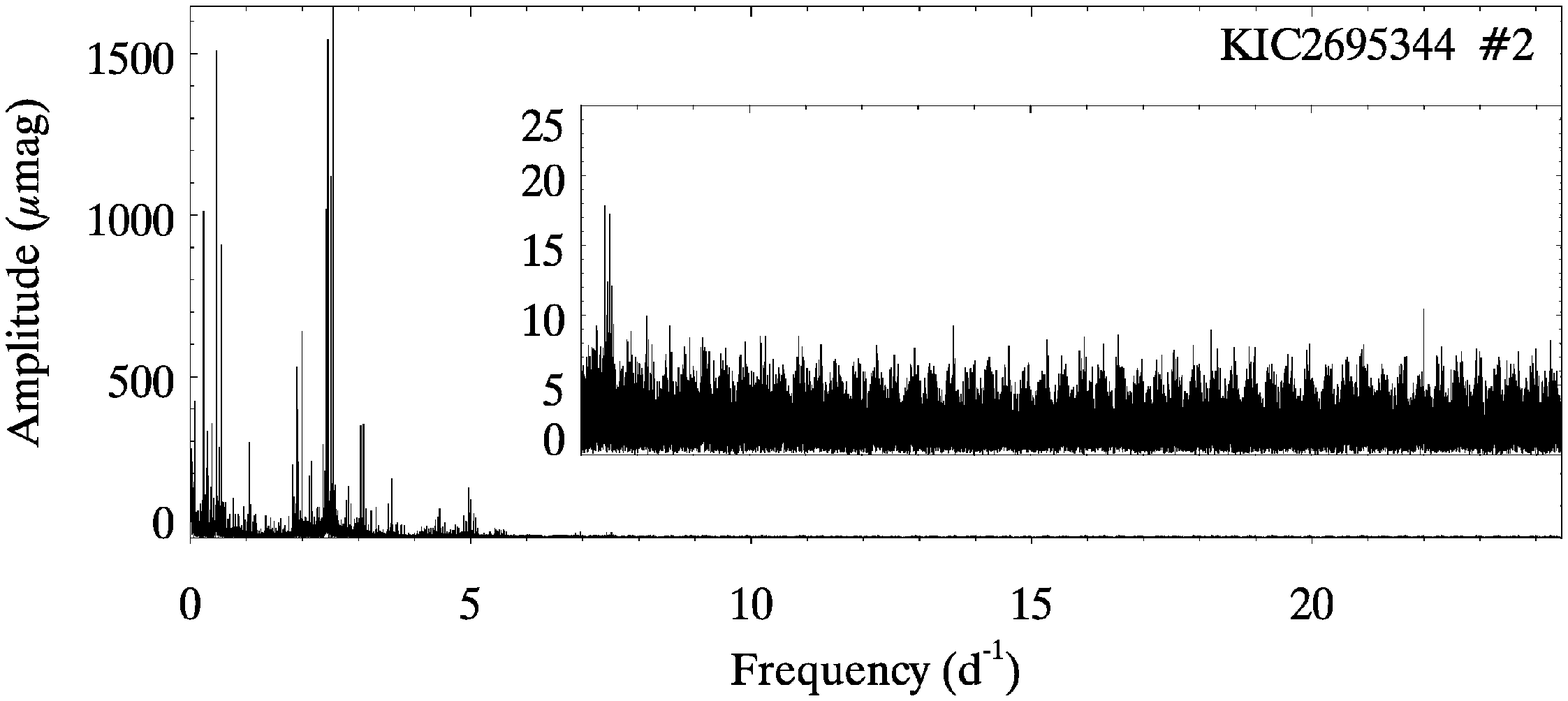}
\includegraphics[width=0.38\textwidth]{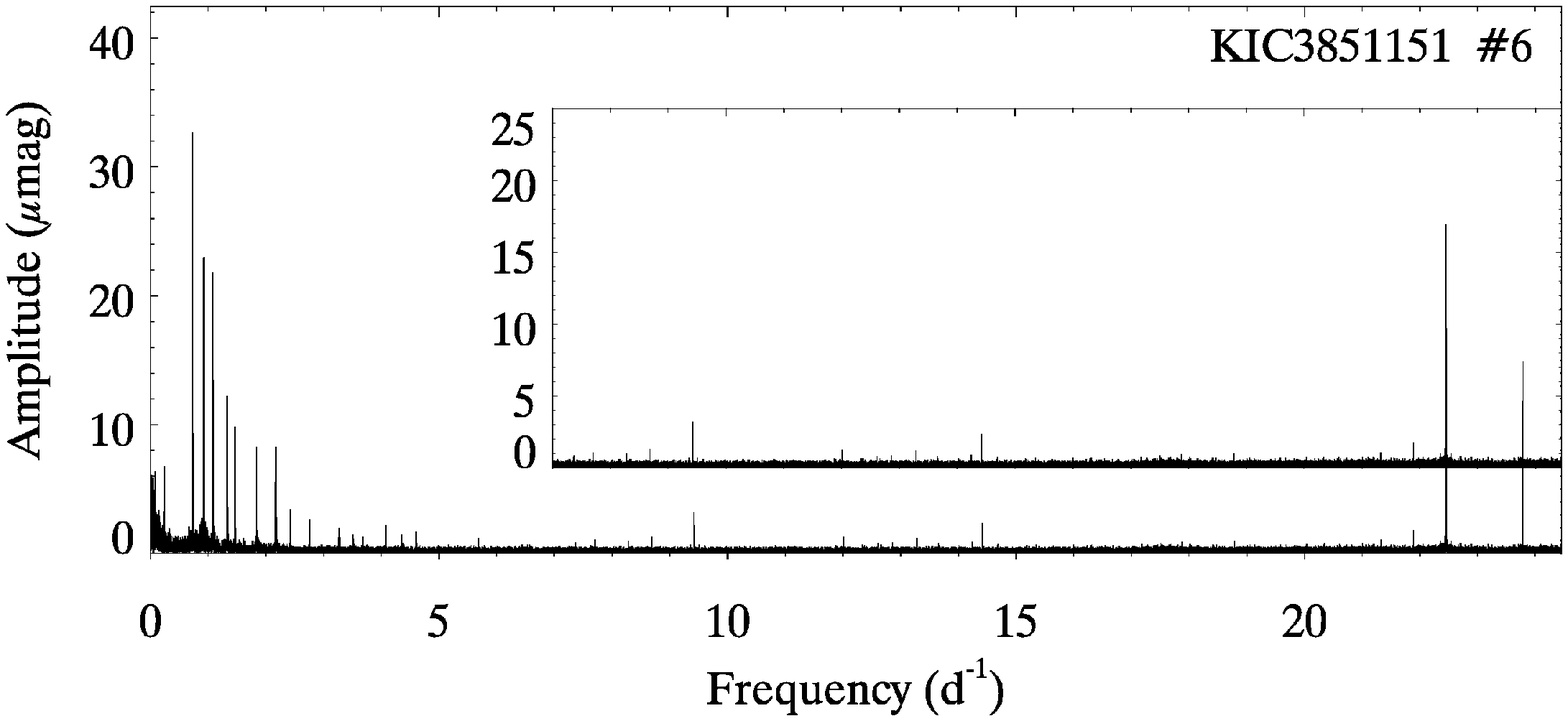}
\includegraphics[width=0.38\textwidth]{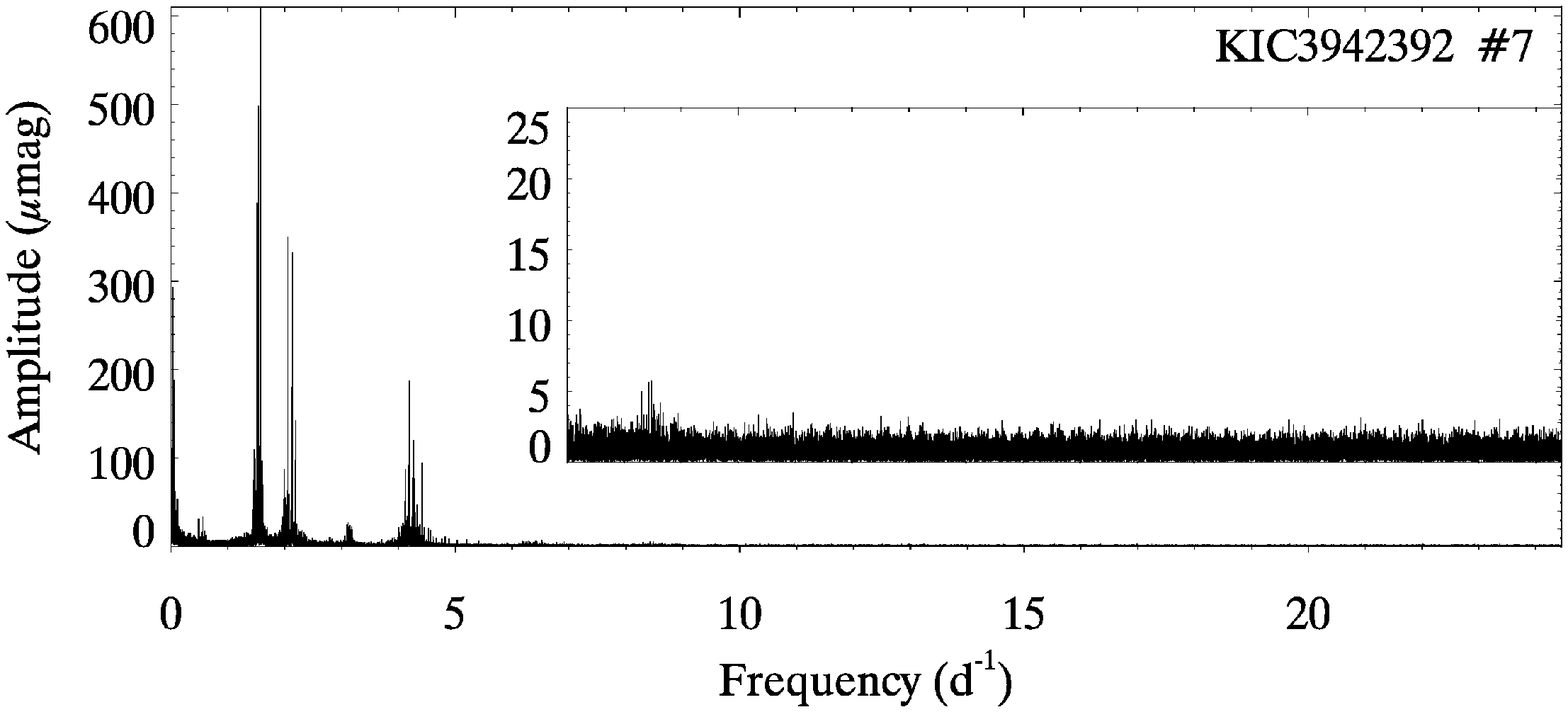}
\includegraphics[width=0.38\textwidth]{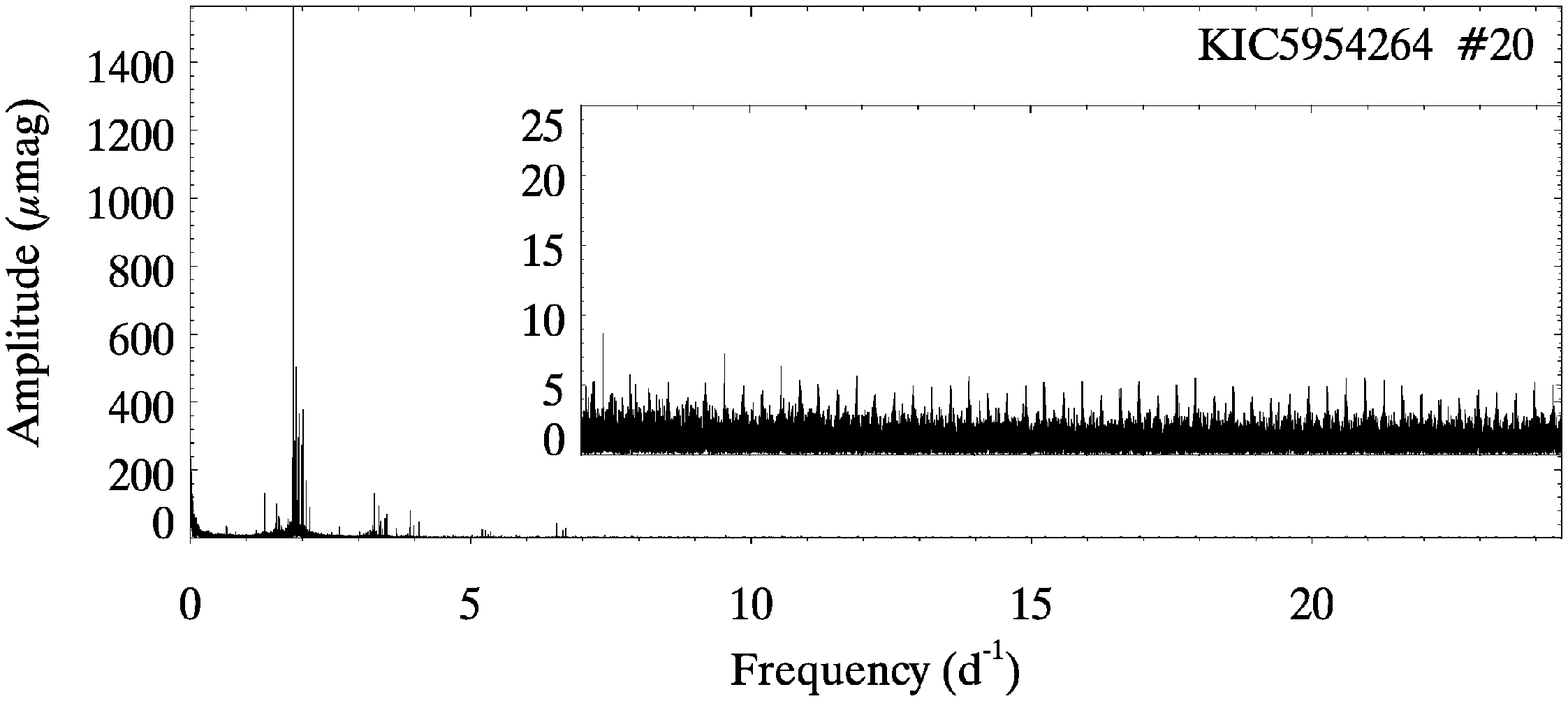}
\includegraphics[width=0.38\textwidth]{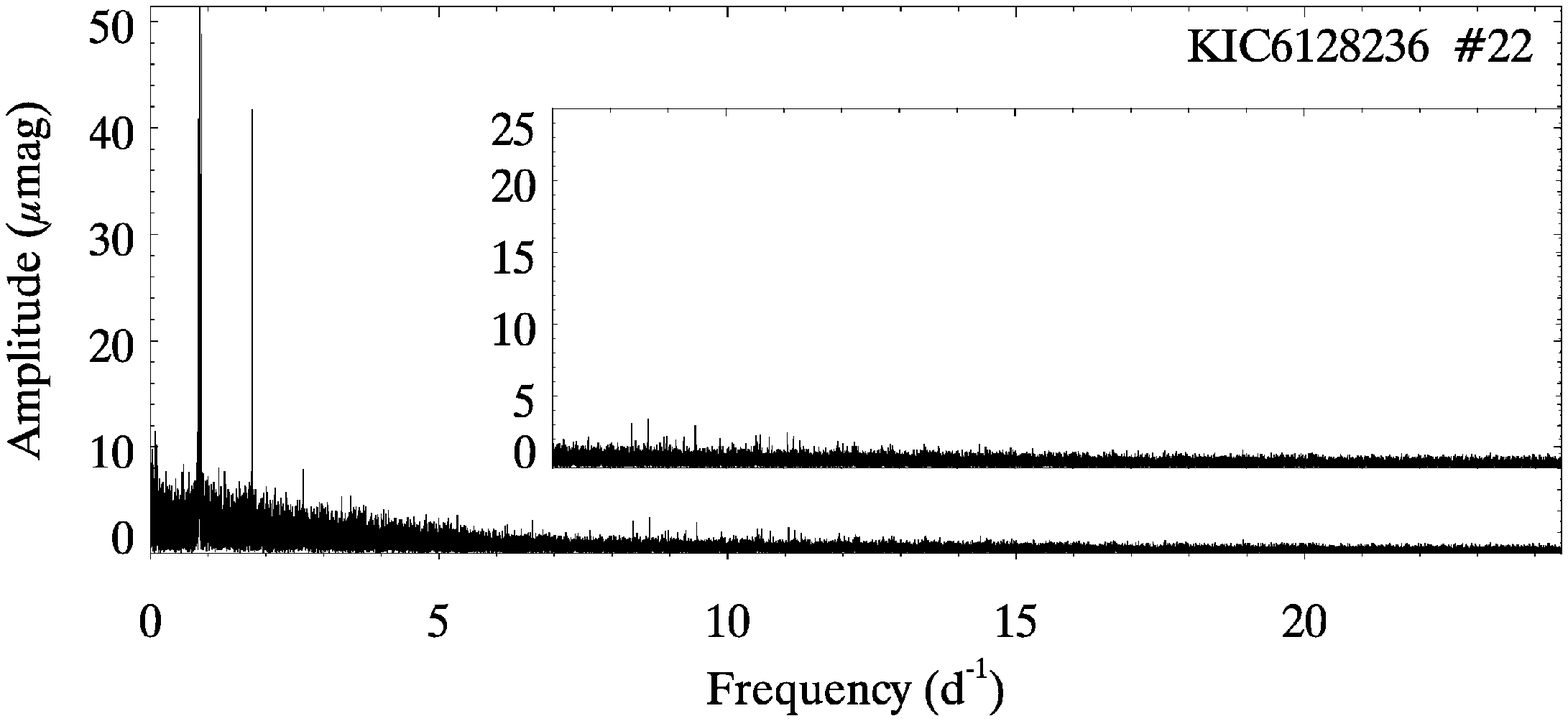}
\includegraphics[width=0.38\textwidth]{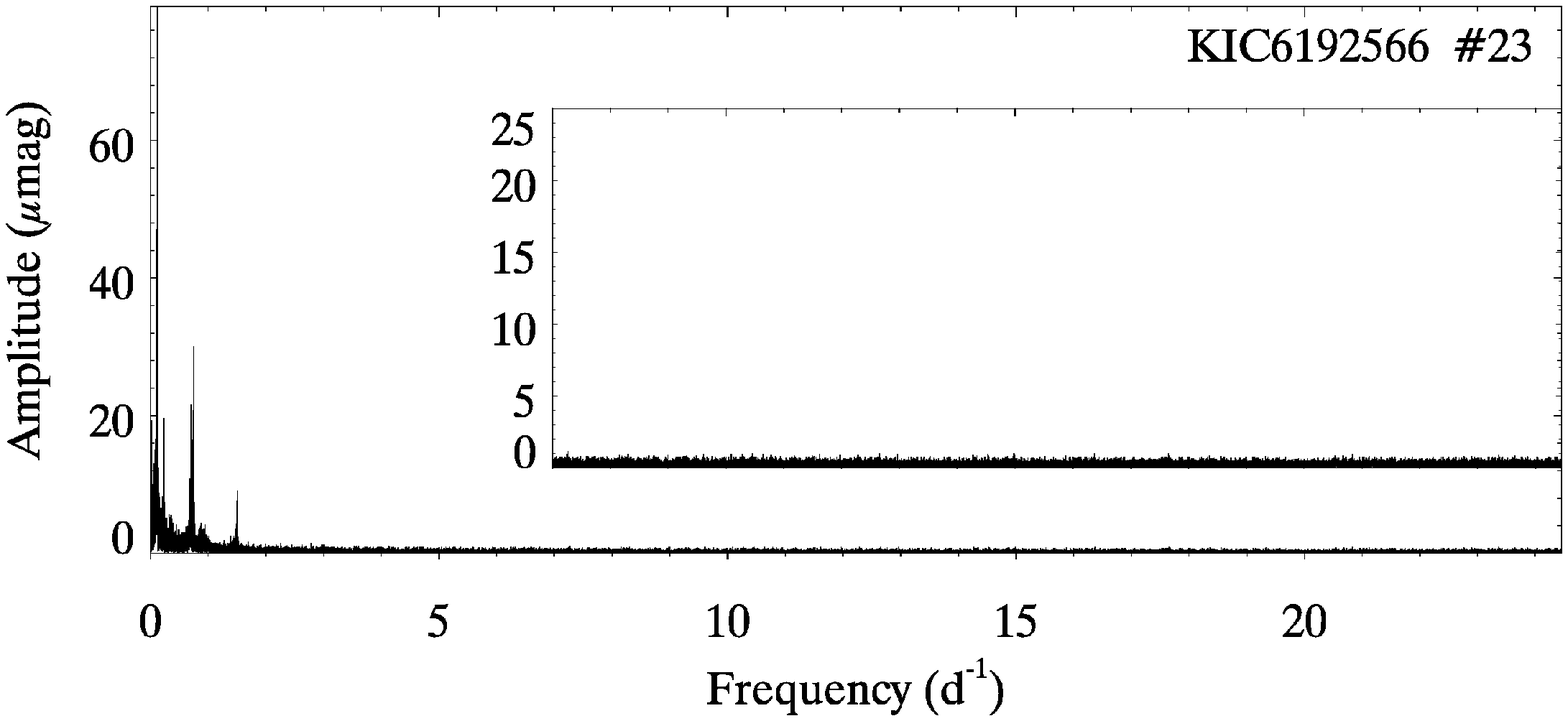}
\includegraphics[width=0.38\textwidth]{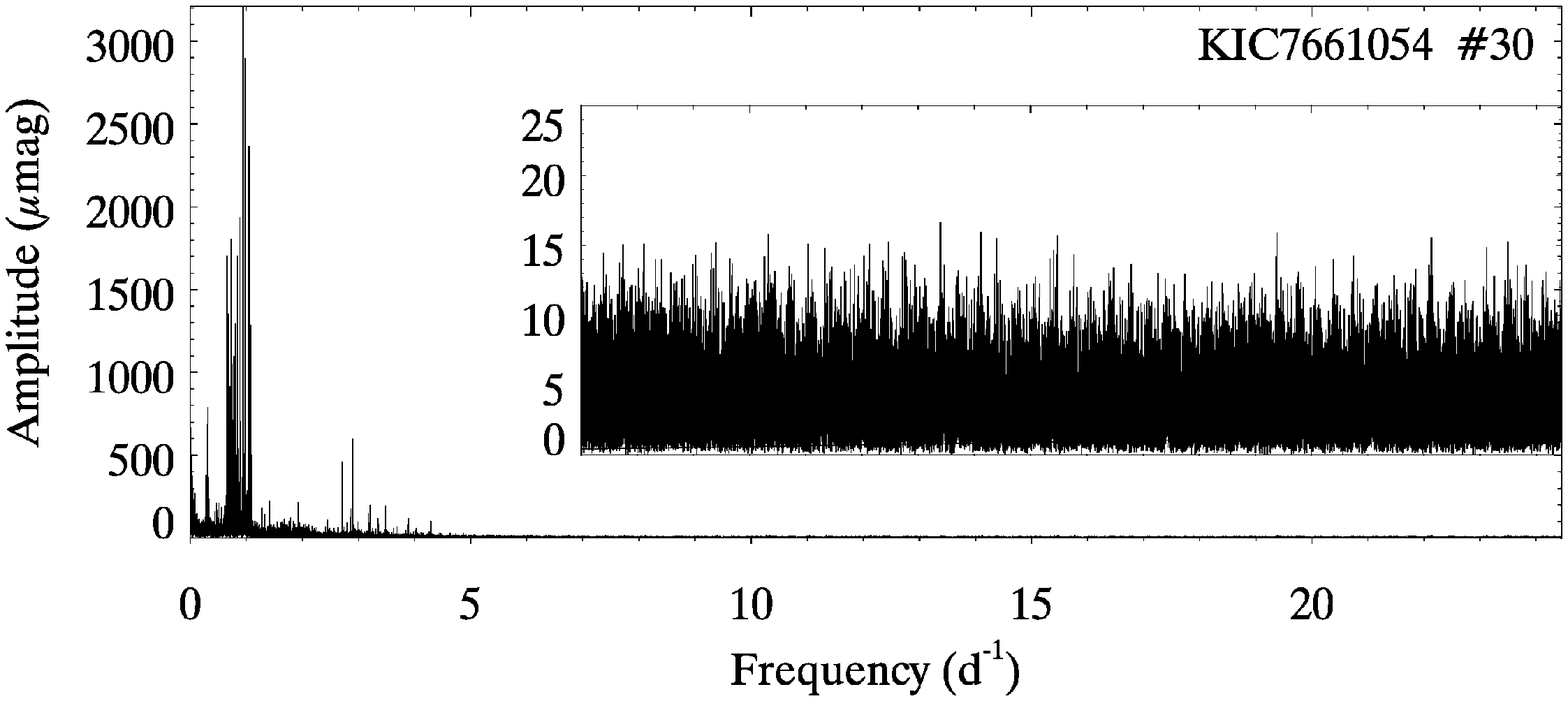}
\includegraphics[width=0.38\textwidth]{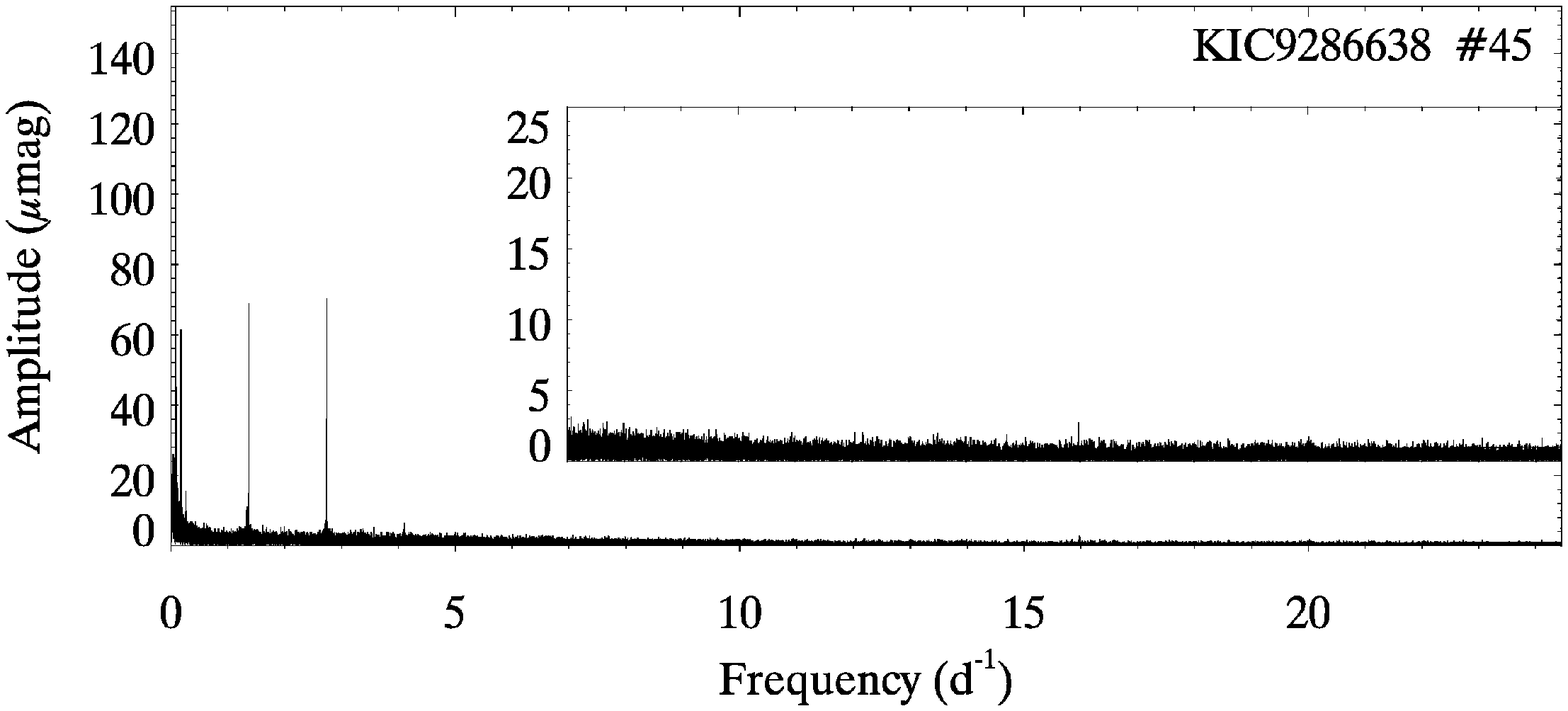}
\includegraphics[width=0.38\textwidth]{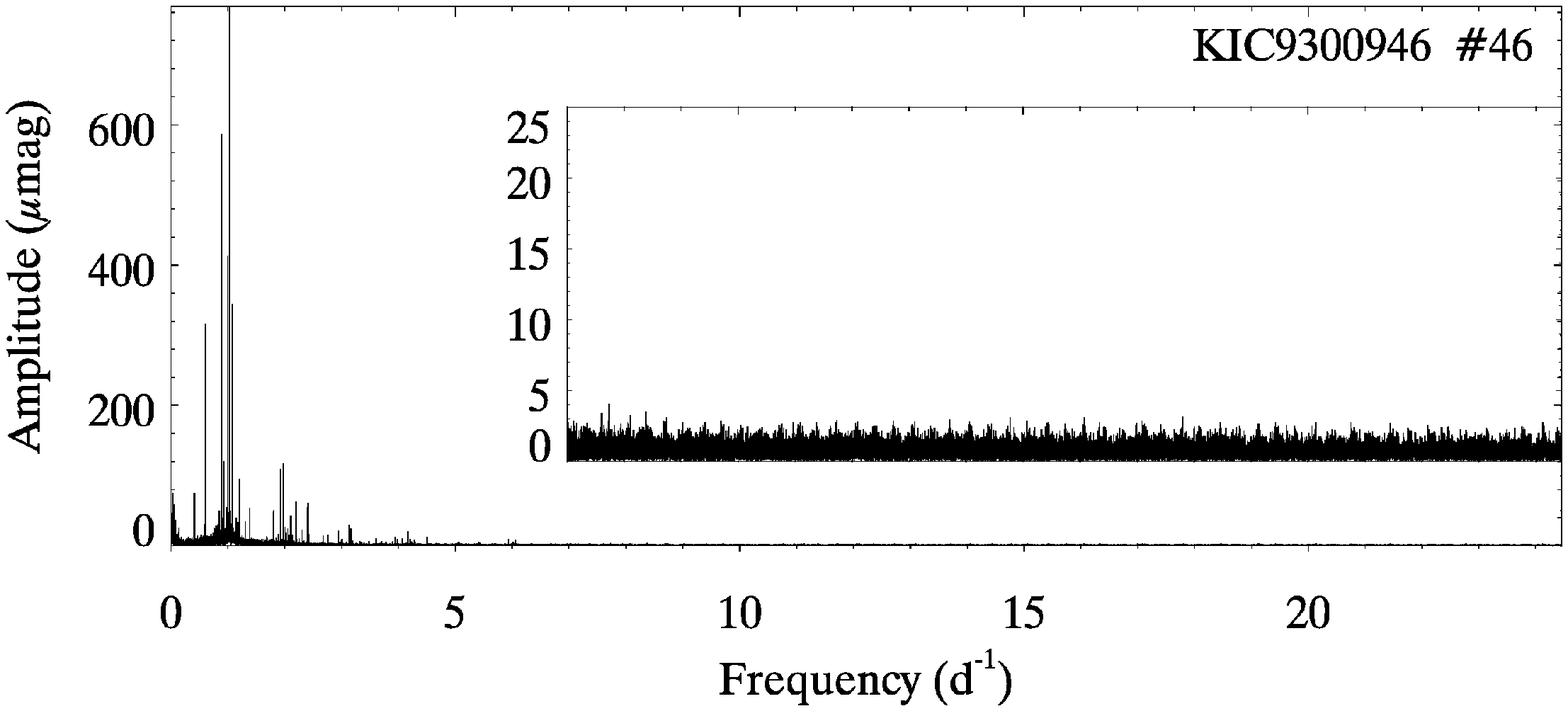}
\includegraphics[width=0.38\textwidth]{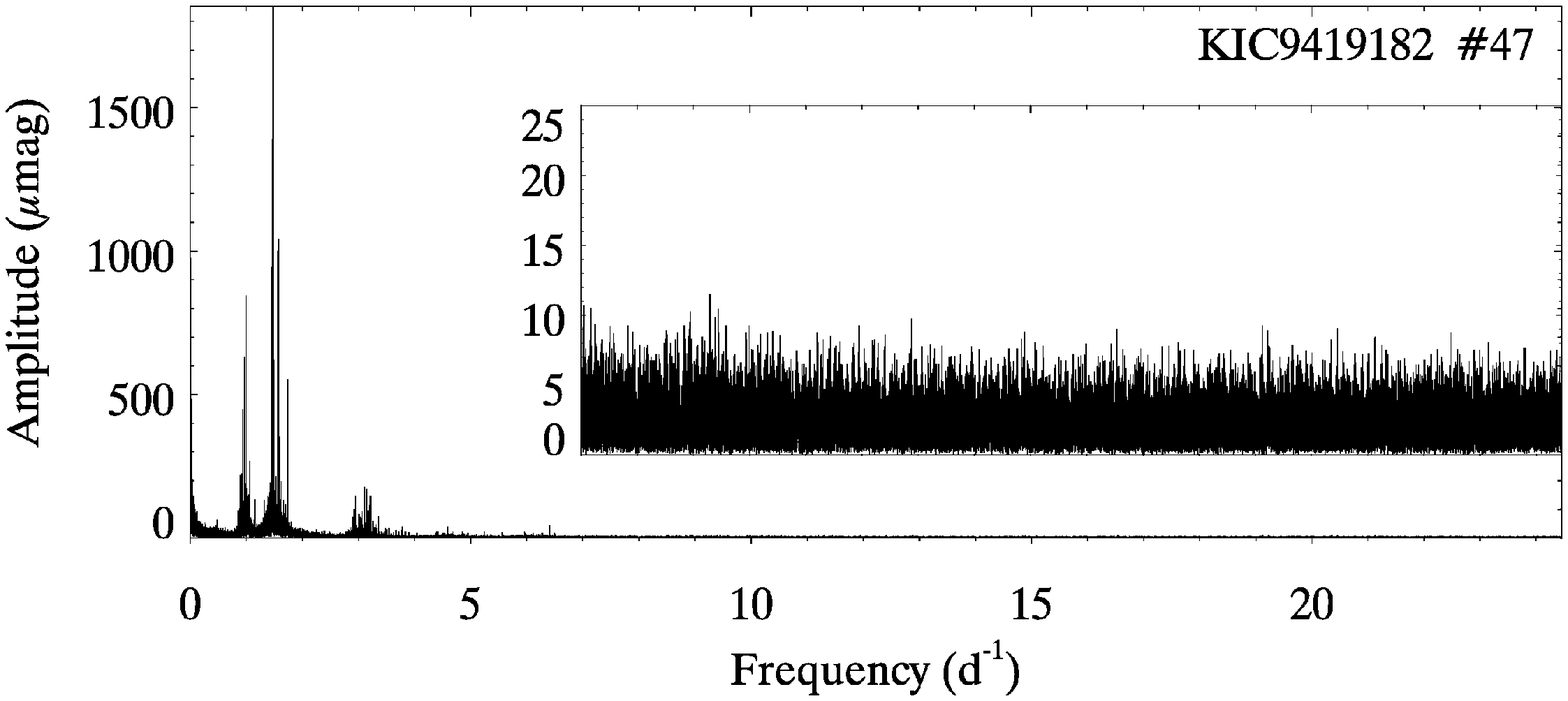}
\includegraphics[width=0.38\textwidth]{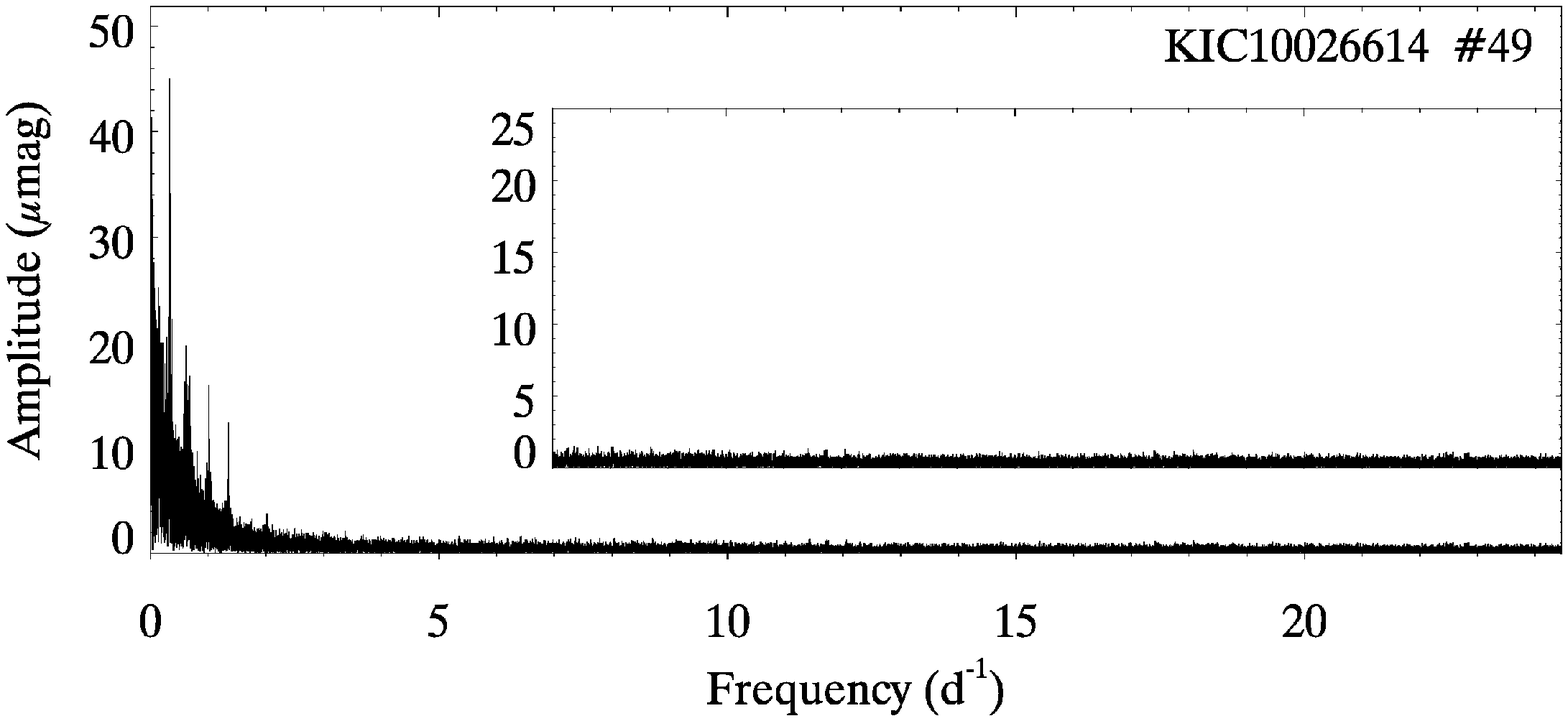}
\includegraphics[width=0.38\textwidth]{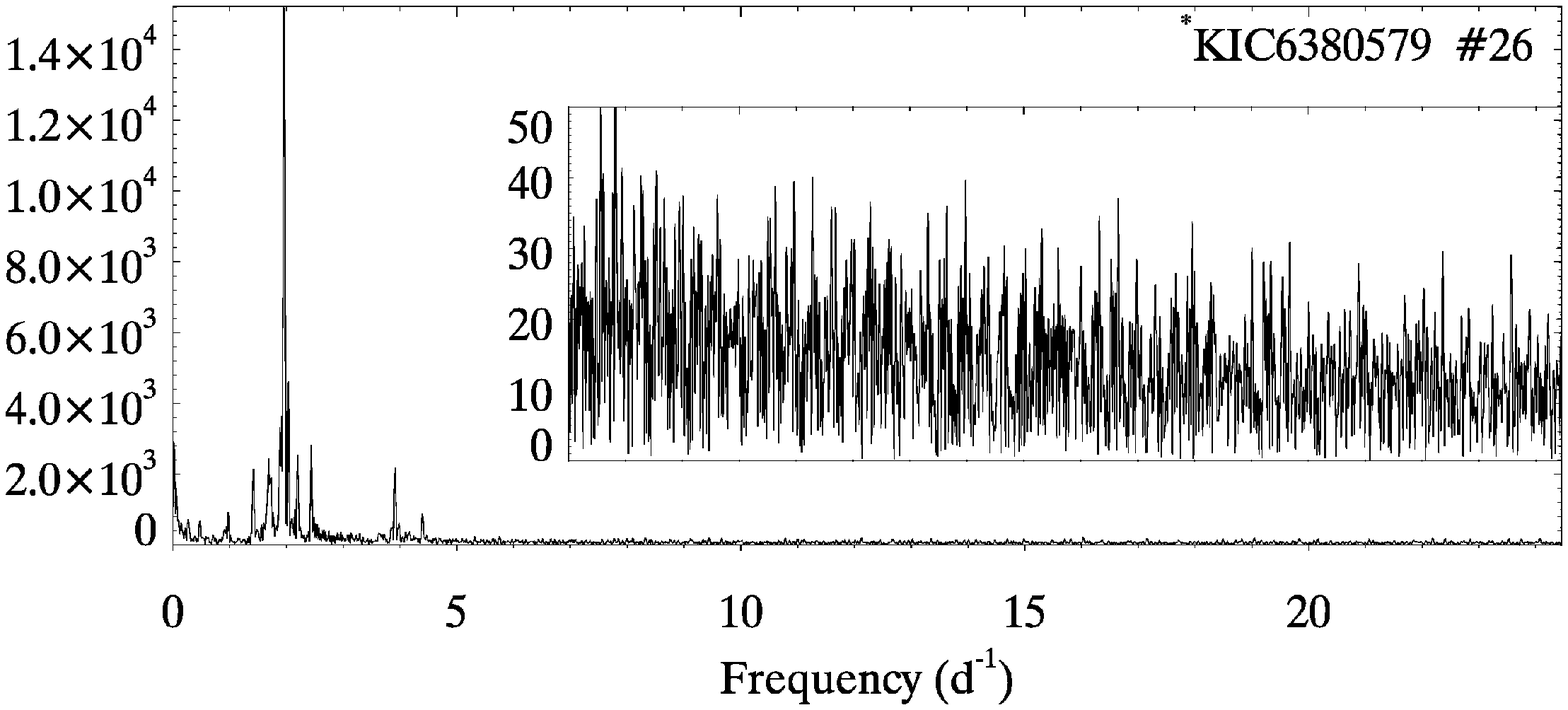}
\caption{Fourier transforms of the light curves of the thirteen stars in Table\:\ref{tab:selected}. Star\# is written in the {\it upper-right corner}. The p-mode region is shown on a zoomed y-scale on the insets, using the same x-scale (7--24.4\,d$^{-1}$). Clearly some stars are $\gamma$\,Dor stars, but not $\delta$\,Sct stars. For one star (\#26, {\it bottom panel}), the g-mode amplitudes are so high we prewhitened the 19 highest peaks for the p-mode inset. Star \#6 shows low-amplitude p\:modes, but with amplitudes below 20\,$\upmu$mag. It is thus below our 50\,$\upmu$mag limit, and the discussion in the text is still relevant to this star.}
\label{fig:fts}
\end{center}
\end{figure*}

Let us suppose that a close binary does not produce an Am star for some reason, and might masquerade as a non-$\delta$\,Sct star inside the instability strip in our sample. One way we might detect such a system is in the \kepler\ light curves, even if the stars do not eclipse each other, from some or all of: tidal distortion, heating, and Doppler beaming \citep{faigler&mazeh2011,tal-oretal2014}. They usually produce a Fourier series of sharp peaks at multiples of the orbital frequency. We provide Fourier transforms of the light curves of each star in Fig.\,\ref{fig:fts}, which we inspected for binary signatures. No additional binaries were detected; low frequency variations in a few of the stars (e.g. \#22, \#23, \#26 and \#45) were temporally variant and gave broad peaks, perhaps arising from rotation instead \citep{balona2014}. We conclude that close binarity has not led to grossly incorrect temperatures for stars in our sample.

We return our discussion to the one chemically normal star in our sample (\#22 = KIC\,612823 = HD\,184521) that does not appear to be a binary star, does not pulsate, but does lie in the middle of the $\delta$\,Sct instability strip, albeit in the hydrogen shell burning phase. Its temperature according to spectroscopy, SED fitting and revised KIC photometry \citep{huberetal2014} are all in agreement to 100\,K, and place it 500\,K from the red edge of the instability strip. Its chemical abundance pattern is normal, hence this is the first confirmed chemically normal, non-pulsator lying inside the $\delta$\,Sct instability strip, and is worthy of further study.

\subsection{$\gamma$\,Dor stars}

Seven of the Fourier transforms in Fig.\,\ref{fig:fts} show obvious g-mode oscillations (Stars \#2, 20, 7, 30, 46, 47, 26). These are $\gamma$\,Dor stars, whose oscillations are driven by convective flux blocking \citep{guziketal2000b,dupretetal2004}. One that does not is Star \#23, which is too hot to be a $\gamma$\,Dor star. Stars \#6 and \#45 have peaks in the frequency range where g\:modes are typically observed, but the peaks are not typical of $\gamma$\,Dor stars (they are not consecutive overtones of g\:modes). Finally, Star \#6 is far beyond the blue edge of the $\gamma$\,Dor instability strip, so the presence of an irregular series of peaks below 5\,d$^{-1}$ cannot be explained, unless this is a binary consisting of a $\gamma$\,Dor star and a $\delta$\,Sct star in their respective instability strips, or the convective blocking mechanism continues to operate in hot A stars \citep{balona2014}. Our observation that the $\gamma$\,Dor pulsators all appear to lie inside the $\gamma$\,Dor instability strip is in disagreement with \citet{uytterhoevenetal2011}, who showed that KIC photometry places some $\gamma$\,Dor stars well outside the $\gamma$\,Dor instability strip boundaries. Rather, our observations agree with \citet{tkachenkoetal2013} who constrained the $T_{\rm eff}$ of 36 $\gamma$\,Dor stars to lie within the $\gamma$\,Dor instability strip with high-resolution spectra. Clearly, high precision on $T_{\rm eff}$ is required to place stars with respect to this (narrow) instability strip.

\subsection{Explaining the lack of p\:modes}

Various mechanisms for limiting pulsation amplitudes exist that might explain the lack of pulsation in our selected stars. One is convection, as we have already mentioned. This probably explains the lack of pulsation in most of the cooler stars in our sample, since they lie within 2$\sigma$ of the red edge of the instability strip. Saturation of the pulsational driving is another. This phenomenon may limit amplitudes in high-amplitude $\delta$\,Sct stars (HADS stars), but it is not a viable explanation for the non-pulsators. A third mechanism is mode coupling. In this scenario, energy is transferred between modes in a parametric resonance \citep{dziembowski&krolikowska1985}, and it is of particular importance that this can occur between p and g\:modes \citep{buchleretal1997}. Seven of the stars in question are g\:mode pulsators. Could it be that the observable amplitudes of the g\:modes, which exist predominantly in the deep interior and do not attain large amplitudes near the surface, result from a transfer of energy from the p\:modes? We find this to be unlikely for two reasons: (i) it cannot explain the lack of p\:modes for the stars that show no g\:mode oscillations, and (ii) $\delta$\,Sct--$\gamma$\,Dor hybrids are common in the \kepler\ data, covering a whole range of pulsation amplitude ratios between the p and the g\:modes \citep{grigahceneetal2010b, uytterhoevenetal2011}.

We also considered whether the non-$\delta$\,Sct stars might be pulsating in high-degree modes. Modes of higher degrees have diminishing amplitudes due to partial cancellation of nodes and antinodes when integrating over the stellar disk (see \citealt{aertsetal2010}, figure\:1.5). If the non-$\delta$\,Sct stars in the instability strip were pulsating exclusively in high-degree modes, it is possible that their amplitudes as observed by \kepler\ would be below the micro-magnitude noise-level. For stars cooler than 7500\,K, there is a substantial photospheric contribution to the noise level, while hotter A stars appear to be limited by instrumental noise \citep{murphy2014}. Although longer observations could reduce the noise level, \kepler\ is no longer pointing at the original field, and observations similar in precision, duration and continuity are not currently feasible. In addition, we find the high-degree modes hypothesis unlikely because there is no known selection mechanism to force high-degree modes and suppress low-degree ones.

Finally, we considered that binarity itself may inhibit the pulsations. Tidally-induced pulsations are well-studied, not only for p\:modes (e.g. \citealt{hambletonetal2013}), but also g\:modes in main sequence stars \citep{fuller&lai2012} and white dwarfs \citep{fuller&lai2011}. However, tidal damping is also known to occur: \citet{waelkens&rufener1983b} looked at stars that fall within the $\beta$\,Cep instability strip, and found that those in close binaries did not pulsate. Additionally, \citet{hambletonetal2013b} found that around 20\:per\:cent of `heartbeat stars' -- eccentric binaries with significant tidal distortion at periastron -- have tidally induced pulsation, but also found some heartbeat stars in the $\gamma$\,Dor instability strip with no self-excited g\:modes, i.e. systems where binarity is damping the high-order g-mode pulsation. 

It has recently been established that solar-like oscillations can be suppressed in close binary systems: \citet{springer&shaviv2013} found that tidally distorted envelopes, where the density is low, cause rapid dissipation of acoustic waves. In addition, \citet{gaulmeetal2014} established that the effect of close binarity on stellar rotation velocity can lead to generation of substantial dynamo magnetic fields, which suppress p\:modes. Both of these mechanisms of tidal damping operate only at very close separation of a few stellar radii. For A stars, separations of a few stellar radii manifest clearly in the Fourier transforms of {\it Kepler} light curves. Since we do not observe the signatures of close binarity in our non-pulsators, we infer that these two mechanisms of tidal damping are unimportant. Close binarity in A stars also generates Am peculiarities, and the Am stars were already filtered out of our sample.

In summary, although we cannot rule out that tidal damping plays a role in the suppression of p\:modes in chemically normal non-pulsators, a viable mechanism appears to be lacking. We also considered whether magnetic damping may still be relevant. Vega, the archetypical A star, possesses a magnetic field of $-0.6\pm0.3$\,G \citep{lignieresetal2009}, and it is reasonable to assume chemically normal A stars possess fields of similar strengths, though they would be difficult to detect. \citet{saio2005} found that a dipole field of 1\,kG is required to stabilise low-order p\:modes, hence magnetic damping is unimportant for all but the strongly magnetic Ap stars.

\citet{guziketal2014} have also studied non-pulsating stars in the $\delta$\,Sct and $\gamma$\,Dor instability strips. Using atmospheric parameters from the original KIC they found 5 non-pulsating stars in the $\delta$\,Sct instability strip. The remainder of their 633-star sample were either pulsators (40\:per\:cent) or had temperatures cooler than the red edge of the $\delta$\,Sct instability strip. The five non-pulsators in the $\delta$\,Sct instability strip are very faint, with the brightest having $Kp = 14.6$\,mag, so high-resolution spectroscopic follow-up is impractical and it is not known whether the stars are chemically peculiar.


\section{Conclusions}

We have examined the distribution of chemically normal, non-pulsating stars in the $\delta$\,Sct instability strip, and found most of them to lie within 200\,K of either the blue or red edge. Given typical uncertainties of 100\,K, we cannot confidently assert that these stars lie within the instability strip.

We investigated whether undetected binary systems could cause temperature discrepancies when interpreted as single stars (Appendix\,\ref{ap:binaries}). Their spectra would give the appearance of a single star inside the instability strip when in fact they form a composite spectrum of two stars that lie outside the instability strip, and do not pulsate for this reason. The temperature discrepancy that can be introduced depends on the orbital phase: a 300-K discrepancy is easily obtained at conjunction for moderate rotators. The spectra of rapid rotators can be difficult to normalise correctly, resulting in another `free parameter' that can make binary systems difficult to detect.

We did find one chemically normal star that does not appear to be in a binary system, does not pulsate, and lies in the middle of the $\delta$\,Sct instability strip, albeit in the post-main sequence evolutionary phase. This star is the first of its kind. We considered multiple mechanisms that might inhibit pulsation in this star, including convective damping, saturation of the pulsational driving mechanism, mode coupling to g\:modes that are only visible in the interior, pulsation exclusively in high-degree modes whose amplitudes diminish when integrated over the entire stellar disk, magnetic damping and tidal damping. None of these mechanisms gives a satisfactory explanation.

While this individual star is a challenge to pulsation theory, the presence of only one chemically normal non-pulsator lying conclusively inside the $\delta$\,Sct instability strip indicates these stars are rare. The essentially null result of \citet{guziketal2014} confirms this fact. Indeed, if the exceptional star can be disqualified in a more detailed analysis, it may be that no chemically normal, non-pulsator occupies this region of the HR diagram.

We conclude by drawing a parallel with the pulsating DAV white dwarfs (DAVs or ZZ\,Ceti stars). Efforts to define the theoretical edges of the DAV instability strip required abandonment of the old `frozen convection' models because of the rapid response of the convection zone to the pulsational perturbation \citep{brickhill1991}. However, convective damping then led to an observed red edge that was hotter than the theoretical one, and to the suspicion that DAVs might pulsate beyond the observed red edge, at amplitudes lower than could be detected \citep[see][for a review]{kotaketal2002}. \citet{mukadametal2004} found the DAV instability strip to be impure, when they used SDSS data as a homogeneous source of atmospheric parameters and found non-pulsators inside instability strip. \citet{kepler2007} later attributed this to the inaccuracy of those data, and said ``The question of the purity of the ZZ\,Ceti instability strip also depends on the accuracy of determination of the effective temperatures and gravities, as the instability strip ranges only around 1200\,K in $T_{\rm eff}$ and depends on gravity.'' Detailed investigation in high-resolution spectroscopic studies \citep{bergeronetal1995,bergeronetal2004,gianninasetal2005,gianninasetal2006} showed the DAV instability strip to be pure, i.e. all stars within the instability strip pulsate and those outside it do not.

The paucity of stars in the $\delta$\,Sct instability strip without p\:modes brings us to the same conclusion about these stars: the $\delta$\,Sct instability strip is pure, unless pulsation is shut down by diffusion or another mechanism, which could be interactions with a companion star. 


\bibliography{8ver-arxiv_non-pulsators} 

\section*{Acknowledgements}

This research was supported by the Australian Research Council. Funding for the Stellar Astrophysics Centre is provided by the Danish National Research Foundation (grant agreement no.: DNRF106). The research is supported by the ASTERISK project (ASTERoseismic Investigations with SONG and Kepler) funded by the European Research Council (grant agreement no.: 267864). E.N. acknowledges support from NCN grant 2011/01/B/ST9/05448. We are grateful to the entire Kepler team for such exquisite data. This research has made use of the SIMBAD database, operated at CDS, Strasbourg, France.

SJM would like to thank the SIfA asteroseismology group plus Hideyuki Saio and Jim Fuller for useful discussions. This work is a continuation of a preliminary analysis by SJM in his PhD thesis.

\appendix

\section{Undetected binaries}
\label{ap:binaries}

We evaluated whether non-pulsators in the $\delta$\,Sct instability strip can be explained as unresolved binaries. If the two components of the binary lie outside the $\delta$\,Sct instability strip, but their composite spectrum is analysed as a single star, it may appear as a non-pulsator inside the instability strip.

The scenario most likely to be observed is a pair of main-sequence stars with a mass ratio close to 1.0. We considered a binary with one star on the blue edge of the $\delta$\,Sct instability strip ($T_{\rm eff} = 8400$\,K and $\log g = 4.0$) and another of the same age at the red edge ($T_{\rm eff} = 6700$\,K and $\log g = 4.2$). The stars have masses of 2.06 and 1.40\,M$_{\odot}$ and luminosities of 25.2 and 4.3\,L$_{\odot}$, respectively. Both have [Fe/H]=0.0. We computed their synthetic spectra using {\sc spectrum} and its auxiliary programs \citep{gray1999}, from {\small ATLAS9} atmospheric models \citep{kurucz1993a} using line lists and opacity distribution functions from \citep{castelli&hubrig2004}. The spectra were rotationally broadened to 100\,km\,s$^{-1}$ using a limb-darkening coefficient of 0.6. This rotation velocity is common in late A stars, for which the modal rotation velocity is 160\,km\,s$^{-1}$ \citep{abt&morrell1995}, but 100\,km\,s$^{-1}$ was chosen to coincide with the measured $v\sin i$ values of apparently single stars in Table\:\ref{tab:selected}. We set the binary orbital period to 20\,d, that being a rough limit below which binaries experience strong tides and become Am stars \citep{debernardi2000}. In the examples that follow, the orbital phase was changed from $0$ to $\upi/2$, corresponding to zero and maximal radial velocity separation of the two components. We combined the spectra using the stellar luminosities as weights. We degraded the composite spectrum to a signal-to-noise ratio (S/N) of 70 to match our HERMES spectra in the equivalent (blue-violet) spectral region.

\begin{figure*}
\begin{center}
\includegraphics[width=0.9\textwidth]{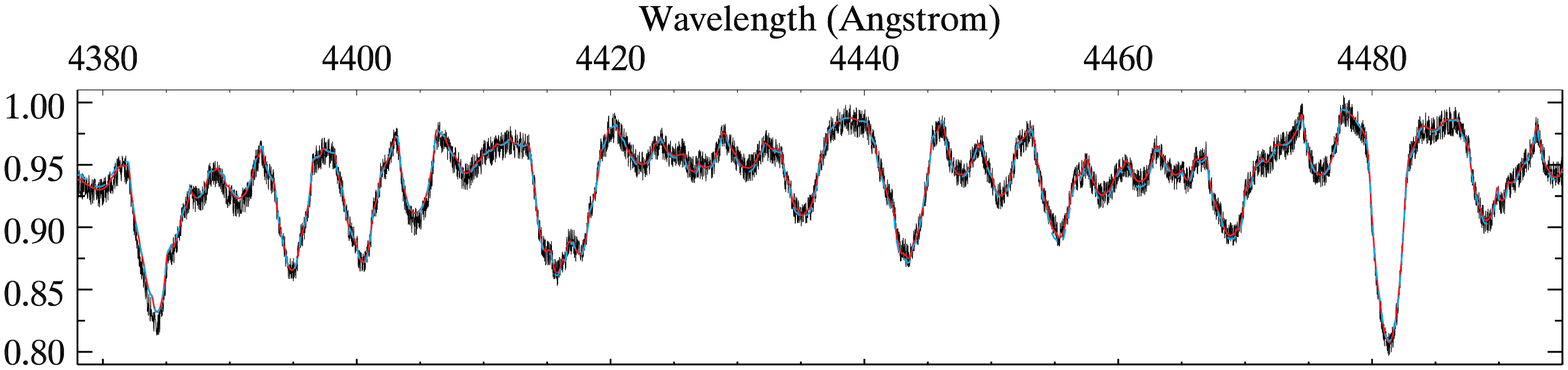}
\includegraphics[width=0.6\textwidth]{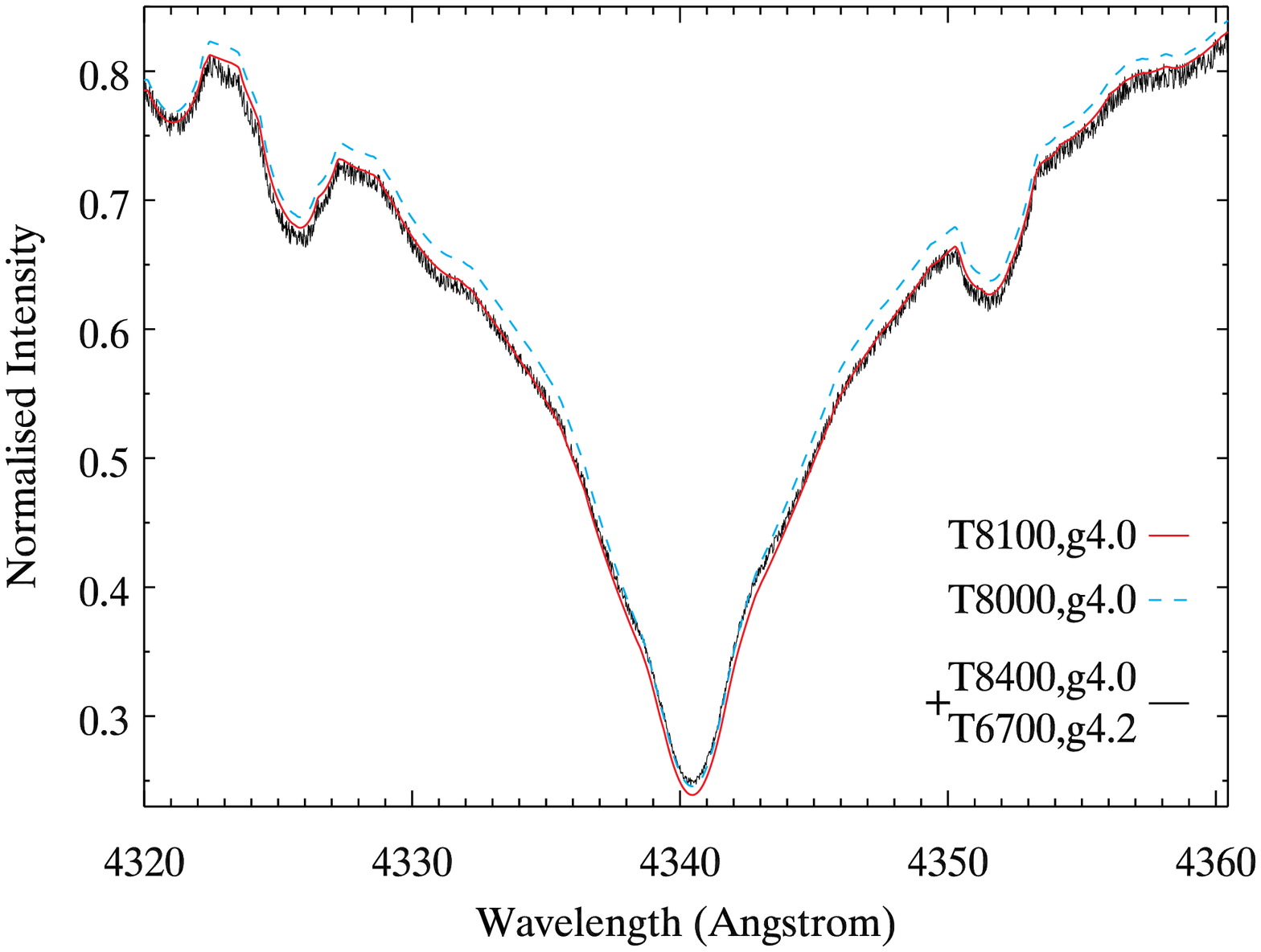}
\includegraphics[width=0.3\textwidth]{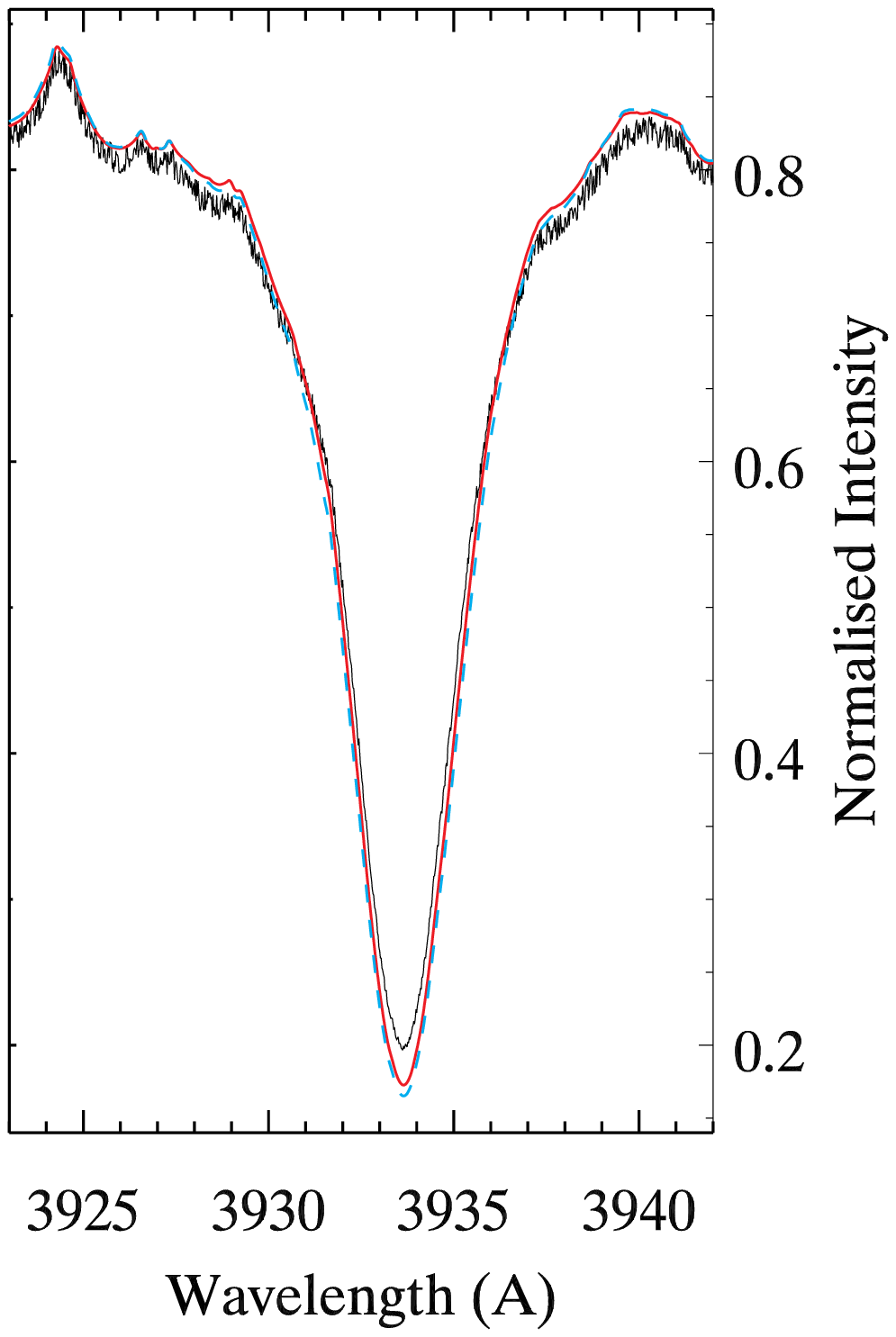}
\caption{Three synthesized spectra: a binary ({\it black lines}) composed of stars at 8400\,K and 6700\,K, having $\log g = 4.0$ and $4.2$, respectively; a single star with $T_{\rm eff} = 8100$\,K and $\log g = 4.0$ ({\it solid red lines}); and a single star with $T_{\rm eff} = 8000$\,K and $\log g = 4.0$ ({\it dashed cyan lines}). The {\it top panel} shows a spectral region rich in metal lines, the {\it left panel} shows the H\,$\gamma$ line and the {\it right panel} shows the \ion{Ca}{ii} K line. Each spectrum has [Fe/H]$=0.0$ and $v \sin i = 100$\,km\,s$^{-1}$. The binary RV separation is 0\,km\,s$^{-1}$, and the spectra are weighted by their luminosity ratio (= 25.2 : 4.3). A $T_{\rm eff}$ discrepancy of 300\,K could be easily introduced if the unresolved binary were treated as a single star.}
\label{fig:binary_phi00}
\end{center}
\end{figure*}

We show the resulting composite spectrum at zero radial-velocity separation in Fig.\,\ref{fig:binary_phi00}, for three spectral regions that are important in the parametrisation of A-star atmospheres: the H\,$\gamma$ line, the \ion{Ca}{ii} K line and a region rich in metal lines at around 4400\,\AA. The wings of the H\,$\gamma$ line have high temperature sensitivity in A stars. In Fig.\,\ref{fig:binary_phi00}, the wings of H\,$\gamma$ for the composite spectrum match a synthetic spectrum of a single star with $T_{\rm eff} = 8100\,$K, i.e.\ 300\,K lower than the primary of the binary system. The fit to the core is not as good, but this is a common problem in high-resolution spectral analyses and is not of concern. An 8000-K star is also shown to illustrate the 100-K (1-$\sigma$) error bars in $T_{\rm eff}$, and is a poorer match to the H\,$\gamma$ wings. Metal lines are not as sensitive to temperature in these stars, so they match both the 8000-K and 8100-K stars. Neither spectrum fits the composite well near the \ion{Ca}{ii} K line. This is the spectral region most likely to hint at a composite spectrum, given its rapid change in strength with temperature. However, at the peak of the Balmer jump, it is also very prone to normalisation errors, which could easily be responsible for such small differences in observed line strength. Furthermore, this resonance line is extremely sensitive to the Ca abundance, which even in normal A stars shows large star-to-star scatter (up to 0.6\,dex; \citealt{gebranetal2010}). From this example we conclude that 300-K discrepancies in $T_{\rm eff}$ are easily introduced when a composite spectrum is interpreted as a single star.

We made a similar comparison (same $v\sin i$ and orbital phase) for a slightly more evolved primary star ($T_{\rm eff} = 8200$\,K, $\log g = 3.8$) of the same mass. The more massive star evolves more rapidly than the 1.4\,M$_{\odot}$ star, whose $\log g$ value is unchanged (to 2 significant figures) at the increased age. Now, the increased luminosity of the primary increases its dominance of the composite spectrum, and the temperature discrepancy is reduced to 150\,K (or 1.5\,$\sigma$).

Binaries composed of a post-main sequence star and a main-sequence star of lower mass are less likely, due to the short timescales for which a post-main sequence star is observed near the $\delta$\,Sct instability strip. However, given the small numbers of non-pulsators observed inside the instability strip, their consideration remains important. Furthermore, if Star \#22 were to be explained as a composite spectrum, the hypothetical binary system would clearly require a bright, red component. We combined spectra of a 2.4- and a 2.1-M$_{\odot}$ star at an age of 686\,Myr, when the primary is in the post-main sequence expansion phase ($T_{\rm eff} = 6200$\,K, $\log g = 3.1$) and the secondary is only half-way through its main sequence lifetime ($T_{\rm eff} = 8500$\,K, $\log g = 4.0$). The composite is easily recognised; the giant has very narrow H-line wings and its cool temperature leads to a broad \ion{Ca}{ii} K line, in direct contrast to its hotter main-sequence companion. While a main-sequence star near F0 could mimic the H lines or metal lines independently, it fails utterly at the blend of H\,$\epsilon$ and \ion{Ca}{ii} H at 3968\,\AA. We thus do not expect composite spectra of post-main sequence and main-sequence stars to be misinterpreted as single stars.

We have heretofore presented examples with orbital phases of zero (without eclipses), where the spectra have zero radial velocity difference. Small radial velocity differences are also obtained for long orbital periods or low-inclination systems. The former cases lend themselves to interferometry and direct imaging, while the latter cases are complicated, as the components are observed with low $v\sin i$ and have non-uniform surface temperatures and gravities due to rotational distortion, as in the case of Vega \citep{petersonetal2006}. Spectroscopy is particularly suited to detecting binary systems where $i \sim 90^{\circ}$ and where substantial radial velocity variation occurs.

\begin{figure*}
\begin{center}
\includegraphics[width=0.9\textwidth]{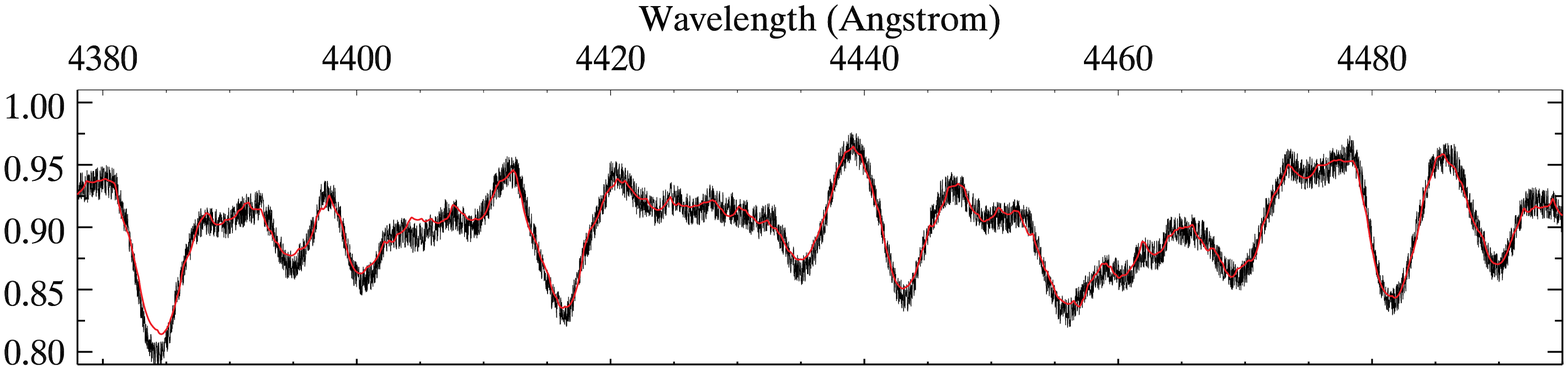}
\includegraphics[width=0.6\textwidth]{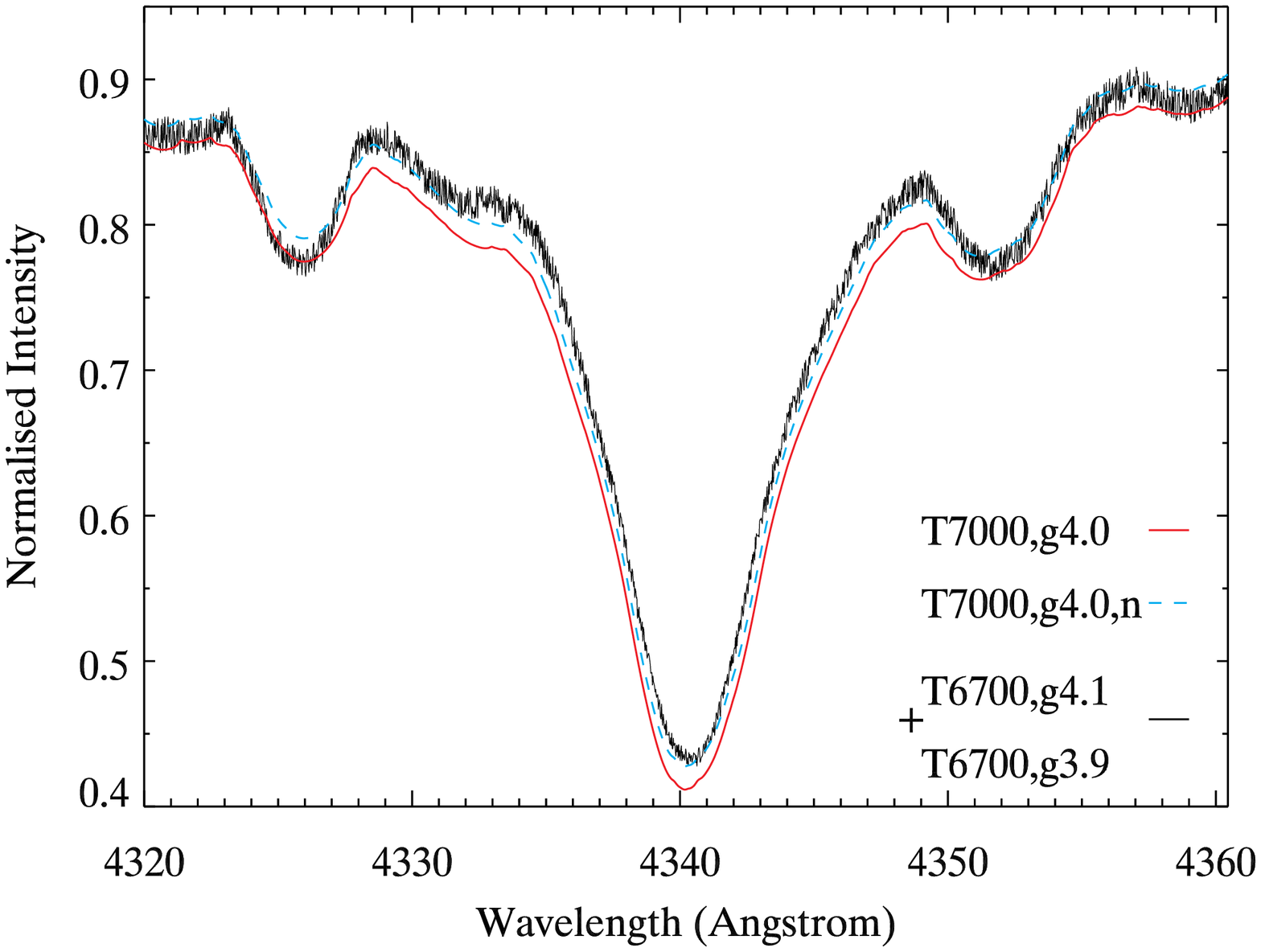}
\includegraphics[width=0.3\textwidth]{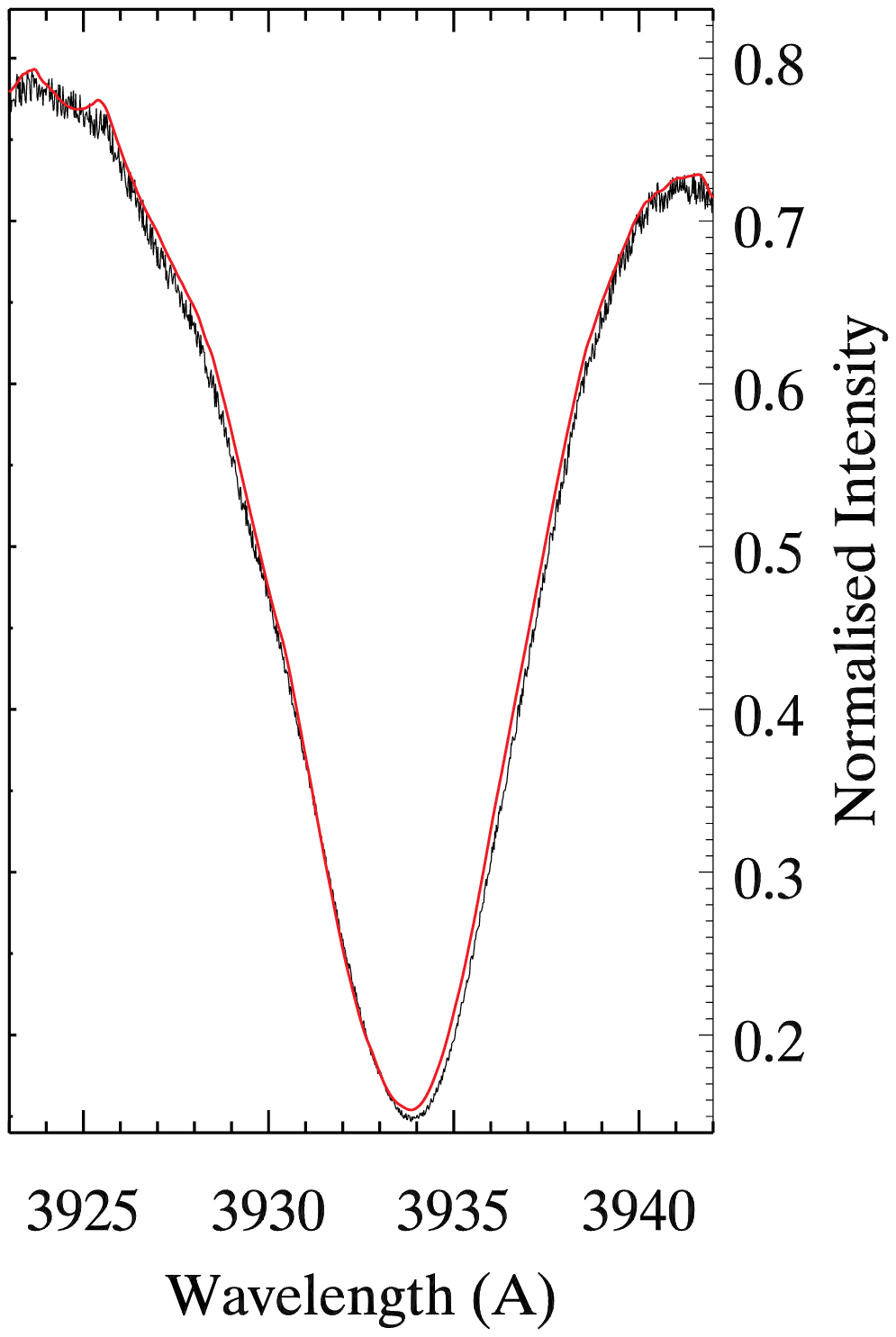}
\caption{A synthetic binary spectrum of stars with $T_{\rm eff} = 6700,6700$\,K, $\log g = 3.9,4.1$, [Fe/H]$ = 0.0,0.0$ and $v\sin i = 160,160$\,km\,s$^{-1}$ ({\it solid black line}). The RV separation is 113\,km\,s$^{-1}$, and the spectral contributions are weighted by luminosity ratio. A single spectrum of a star with $T_{\rm eff} = 7000$\,K, $\log g = 4.0$, [Fe/H]$ = 0.1$ and $v\sin i = 180$\,km\,s$^{-1}$ ({\it solid red line}) is a near morphological match, but a better match with the same spectrum can be achieved with a different normalisation ({\it dashed blue line}, {\it left panel} only).}
\label{fig:binary_phi05}
\end{center}
\end{figure*}

Fig.\,\ref{fig:binary_phi05} shows a composite spectrum at orbital phase $\upi / 2$, where the radial velocity difference amounts to 113\,km\,s$^{-1}$. This is where the stellar rotational velocity becomes important: slow rotators will produce obvious double-lined spectra, while rapid rotators will have blended, asymmetric lines. Finding a line-free continuum for normalisation is an added difficulty in the analysis of rapid rotators, regardless of S/N and spectral resolution. Spectral reduction, including order merging and normalisation, can introduce similar peculiarities into spectra as binarity, and the two effects are difficult to distinguish. The input spectra for Fig.\,\ref{fig:binary_phi05} each have $T_{\rm eff} = 6700$\,K, [Fe/H]$=0.0$ and $v\sin i = 160$\,km\,s$^{-1}$, but the coeval stars have different masses ($m_1 = 1.59$\,M$_{\odot}$, $m_2 = 1.45$\,M$_{\odot}$) and thus different $\log g$ (3.9 and 4.1). We conclude that any of the stars that lie 1 to 2$\sigma$ inside the red edge could result from undetected binaries containing two main-sequence stars just beyond the red edge. This is a natural consequence of the large number of free parameters involved with modelling two stars.

While methods exist for disentangling binary spectra when a set of composite spectra is available for a range of orbital phases \citep[e.g.][]{simon&sturm1994}, the requirements exceed the availability of data in this and indeed most cases, where single-epoch observations are the norm, and where there are no eclipses to `mask' one of the stars.

\pagebreak

\section{Notes on spectral types}
\label{ap:spectraltype}

In this appendix, we provide the notes on spectral types assigned to the stars in Table\:\ref{tab:parameters}. In the following, ALE stands for anomalous luminosity effect (cf.\ \citealt{barry1970}, where the term first appears in print, and also \citealt{gray&corbally2009} for a thorough explanation).

\noindent
\textbf{\#3} Slight Ca weakness in both the K line and $\lambda$4226 line.\\
\textbf{\#9} Metal lines and \ion{Ca}{ii} K line are slightly weaker than A3 in spite of rotation, but hydrogen lines match A3\,Va.\\
\textbf{\#10} Very close to the A1\,IVs standard ($\rho$\,Peg) except for a slightly fainter luminosity class.\\
\textbf{\#14} No obvious weakness in \specline{Ca}{i}{4226}, but \specline{Sr}{ii}{4216} is enhanced. \specline{Sr}{ii}{4077} is clearly enhanced. The redward third of \specline{Mn}{i}{4030} is weakened, as is characteristic of Am stars. Slight ALE.\\
\textbf{\#18} Typical Am star: Ca weak, Sr strong.\\
\textbf{\#19} Rapid rotator. \ion{Ca}{ii} K line best matches A0.5 and is broad. No obvious He lines. Metals weaker than A1. Shallow H cores do not fit at any type.\\
\textbf{\#36} Almost featureless spectrum. Metal lines are absent, with the exception of a very weak \specline{Mg}{ii}{4481} (B7) and weak \ion{Ca}{ii} K line (B8). He lines at $\lambda\lambda$4026 and 4471 are slightly stronger than A0. H wings are only slightly narrower than A0 (indicating B9.5, like the He lines), but are too shallow in the core.\\
\textbf{\#38} Metal line type is about A3 or A4s (assuming luminosity class IV/V), but H line type could not be matched, possibly due to a normalisation issue.\\
\textbf{\#40} Classical Am signature, with enhanced Sr and weak Ca. The redward third of \specline{Mn}{i}{4030} is weak. Luminosity class based on Fe/\specline{Ti}{ii}{4172}-9 blend, as is typical for Am stars, but hydrogen lines match the A3\,Va standard. ALE.\\
\textbf{\#42} Extreme Am star. Sr lines stronger still than the F3 standard. Large ALE.\\
\textbf{\#43} Extreme Am. Slight ALE.\\
\textbf{\#46} A superb match to the F2\,Vs standard, 78\,UMa.\\
\textbf{\#51} Luminosity criteria are in disagreement. The breadth of the H lines favour a dwarf classification, but the \ion{Fe}{ii}/\ion{Ti}{ii} blends favour a giant. This illustrates the difficulty of luminosity classification at A5. The \ion{Ca}{ii} K line type is A6. Compromise type given is A5\,IV, but the star does not match the A5\,IV standard well. A metal-rich, early A ($\sim$A3) star was considered, but neutral metals appear no earlier than A3. Oddly, \specline{Mn}{i}{4030} is enhanced, but is usually expected to show a mild negative luminosity effect.

\end{document}